\documentclass[journal,10pt]{IEEEtran}
\usepackage{cite}
\usepackage{graphicx}
\usepackage{amssymb}
\usepackage{amsmath}
\usepackage{amsthm}
\usepackage{booktabs} 
\usepackage{algorithm}
\usepackage{algorithmic}
\usepackage{microtype}
\usepackage{balance}
\usepackage{grffile} 
\usepackage[dvipsnames]{xcolor}
\usepackage{mathabx}
\usepackage{tikz}
\usetikzlibrary{positioning}
\usepackage{subcaption}
\usepackage[font=footnotesize,labelfont=bf]{caption}
\usepackage{epstopdf}
\usepackage{color, colortbl}
\definecolor{Gray}{gray}{0.9}

\newcommand{\bb}[1]{\mathbb{#1}}
\newcommand{\mb}[1]{\mathbf{#1}}
\newcommand{\mc}[1]{\mathcal{#1}}

\makeatletter
\newcommand*\bigcdot{\mathpalette\bigcdot@{.5}}
\newcommand*\bigcdot@[2]{\mathbin{\vcenter{\hbox{\scalebox{#2}{$\m@th#1\bullet$}}}}}
\makeatother

\usepackage{amssymb}
\usepackage{amsfonts}
\usepackage{mathrsfs}
\usepackage{xspace}
\usepackage{bm}
\usepackage{upgreek}

\newcommand{\safemath}[2]{\newcommand{#1}{\ensuremath{#2}\xspace}}

\safemath{\bma}{\mathbf{a}}
\safemath{\bmb}{\mathbf{b}}
\safemath{\bmc}{\mathbf{c}}
\safemath{\bmd}{\mathbf{d}}
\safemath{\bme}{\mathbf{e}}
\safemath{\bmf}{\mathbf{f}}
\safemath{\bmg}{\mathbf{g}}
\safemath{\bmh}{\mathbf{h}}
\safemath{\bmi}{\mathbf{i}}
\safemath{\bmj}{\mathbf{j}}
\safemath{\bmk}{\mathbf{k}}
\safemath{\bml}{\mathbf{l}}
\safemath{\bmm}{\mathbf{m}}
\safemath{\bmn}{\mathbf{n}}
\safemath{\bmo}{\mathbf{o}}
\safemath{\bmp}{\mathbf{p}}
\safemath{\bmq}{\mathbf{q}}
\safemath{\bmr}{\mathbf{r}}
\safemath{\bms}{\mathbf{s}}
\safemath{\bmt}{\mathbf{t}}
\safemath{\bmu}{\mathbf{u}}
\safemath{\bmv}{\mathbf{v}}
\safemath{\bmw}{\mathbf{w}}
\safemath{\bmx}{\mathbf{x}}
\safemath{\bmy}{\mathbf{y}}
\safemath{\bmz}{\mathbf{z}}
\safemath{\bmzero}{\mathbf{0}}
\safemath{\bmone}{\mathbf{1}}

\bmdefine{\biad}{a}
\bmdefine{\bibd}{b}
\bmdefine{\bicd}{c}
\bmdefine{\bidd}{d}
\bmdefine{\bied}{e}
\bmdefine{\bifd}{f}
\bmdefine{\bigd}{g}
\bmdefine{\bihd}{h}
\bmdefine{\biid}{i}
\bmdefine{\bijd}{j}
\bmdefine{\bikd}{k}
\bmdefine{\bild}{l}
\bmdefine{\bimd}{m}
\bmdefine{\bind}{n}
\bmdefine{\biod}{o}
\bmdefine{\bipd}{p}
\bmdefine{\biqd}{q}
\bmdefine{\bird}{r}
\bmdefine{\bisd}{s}
\bmdefine{\bitd}{t}
\bmdefine{\biud}{u}
\bmdefine{\bivd}{v}
\bmdefine{\biwd}{w}
\bmdefine{\bixd}{x}
\bmdefine{\biyd}{y}
\bmdefine{\bizd}{z}

\bmdefine{\bixid}{\xi}
\bmdefine{\bilambdad}{\lambda}
\bmdefine{\bimud}{\mu}
\bmdefine{\bithetad}{\theta}
\bmdefine{\biphid}{\phi}
\bmdefine{\bideltad}{\delta}

\safemath{\bmia}{\biad}
\safemath{\bmib}{\bibd}
\safemath{\bmic}{\bicd}
\safemath{\bmid}{\bidd}
\safemath{\bmie}{\bied}
\safemath{\bmif}{\bifd}
\safemath{\bmig}{\bigd}
\safemath{\bmih}{\bihd}
\safemath{\bmii}{\biid}
\safemath{\bmij}{\bijd}
\safemath{\bmik}{\bikd}
\safemath{\bmil}{\bild}
\safemath{\bmim}{\bimd}
\safemath{\bmin}{\bind}
\safemath{\bmio}{\biod}
\safemath{\bmip}{\bipd}
\safemath{\bmiq}{\biqd}
\safemath{\bmir}{\bird}
\safemath{\bmis}{\bisd}
\safemath{\bmit}{\bitd}
\safemath{\bmiu}{\biud}
\safemath{\bmiv}{\bivd}
\safemath{\bmiw}{\biwd}
\safemath{\bmix}{\bixd}
\safemath{\bmiy}{\biyd}
\safemath{\bmiz}{\bizd}

\safemath{\bmxi}{\bixid}
\safemath{\bmlambda}{\bilambdad}
\safemath{\bmmu}{\bimud}
\safemath{\bmtheta}{\bithetad}
\safemath{\bmphi}{\biphid}
\safemath{\bmdelta}{\bideltad}

\safemath{\bA}{\mathbf{A}}
\safemath{\bB}{\mathbf{B}}
\safemath{\bC}{\mathbf{C}}
\safemath{\bD}{\mathbf{D}}
\safemath{\bE}{\mathbf{E}}
\safemath{\bF}{\mathbf{F}}
\safemath{\bG}{\mathbf{G}}
\safemath{\bH}{\mathbf{H}}
\safemath{\bI}{\mathbf{I}}
\safemath{\bJ}{\mathbf{J}}
\safemath{\bK}{\mathbf{K}}
\safemath{\bL}{\mathbf{L}}
\safemath{\bM}{\mathbf{M}}
\safemath{\bN}{\mathbf{N}}
\safemath{\bO}{\mathbf{O}}
\safemath{\bP}{\mathbf{P}}
\safemath{\bQ}{\mathbf{Q}}
\safemath{\bR}{\mathbf{R}}
\safemath{\bS}{\mathbf{S}}
\safemath{\bT}{\mathbf{T}}
\safemath{\bU}{\mathbf{U}}
\safemath{\bV}{\mathbf{V}}
\safemath{\bW}{\mathbf{W}}
\safemath{\bX}{\mathbf{X}}
\safemath{\bY}{\mathbf{Y}}
\safemath{\bZ}{\mathbf{Z}}

\safemath{\bZero}{\mathbf{0}}
\safemath{\bOne}{\mathbf{1}}
\safemath{\bDelta}{\mathbf{\Delta}}
\safemath{\bLambda}{\mathbf{\UpLambda}}
\safemath{\bPhi}{\mathbf{\Phi}}
\safemath{\bPsi}{\mathbf{\Psi}}
\safemath{\bSigma}{\mathbf{\Upsigma}}
\safemath{\bOmega}{\mathbf{\Upomega}}
\safemath{\bTheta}{\mathbf{\Uptheta}}

\bmdefine{\biAd}{A}
\bmdefine{\biBd}{B}
\bmdefine{\biCd}{C}
\bmdefine{\biDd}{D}
\bmdefine{\biEd}{E}
\bmdefine{\biFd}{F}
\bmdefine{\biGd}{G}
\bmdefine{\biHd}{H}
\bmdefine{\biId}{I}
\bmdefine{\biJd}{J}
\bmdefine{\biKd}{K}
\bmdefine{\biLd}{L}
\bmdefine{\biMd}{M}
\bmdefine{\biOd}{N}
\bmdefine{\biPd}{O}
\bmdefine{\biQd}{P}
\bmdefine{\biRd}{R}
\bmdefine{\biSd}{S}
\bmdefine{\biTd}{T}
\bmdefine{\biUd}{U}
\bmdefine{\biVd}{V}
\bmdefine{\biWd}{W}
\bmdefine{\biXd}{X}
\bmdefine{\biYd}{Y}
\bmdefine{\biZd}{Z}

\bmdefine{\biDelta}{\Delta}
\bmdefine{\biLambda}{\Lambda}
\bmdefine{\biPhi}{\Phi}
\bmdefine{\biSigma}{\Sigma}
\bmdefine{\biOmega}{\Omega}
\bmdefine{\biTheta}{\Theta}

\safemath{\bimA}{\biAd}
\safemath{\bimB}{\biBd}
\safemath{\bimC}{\biCd}
\safemath{\bimD}{\biDd}
\safemath{\bimE}{\biEd}
\safemath{\bimF}{\biFd}
\safemath{\bimG}{\biGd}
\safemath{\bimH}{\biHd}
\safemath{\bimI}{\biId}
\safemath{\bimJ}{\biJd}
\safemath{\bimK}{\biKd}
\safemath{\bimL}{\biLd}
\safemath{\bimM}{\biMd}
\safemath{\bimN}{\biNd}
\safemath{\bimO}{\biOd}
\safemath{\bimP}{\biPd}
\safemath{\bimQ}{\biQd}
\safemath{\bimR}{\biRd}
\safemath{\bimS}{\biSd}
\safemath{\bimT}{\biTd}
\safemath{\bimU}{\biUd}
\safemath{\bimV}{\biVd}
\safemath{\bimW}{\biWd}
\safemath{\bimX}{\biXd}
\safemath{\bimY}{\biYd}
\safemath{\bimZ}{\biZd}

\safemath{\bimDelta}{\biDelta}
\safemath{\bimLambda}{\biLambda}
\safemath{\bimPhi}{\biPhi}
\safemath{\bimSigma}{\biSigma}
\safemath{\bimOmega}{\biOmega}
\safemath{\bimTheta}{\biTheta}

\safemath{\setA}{\mathcal{A}}
\safemath{\setB}{\mathcal{B}}
\safemath{\setC}{\mathcal{C}}
\safemath{\setD}{\mathcal{D}}
\safemath{\setE}{\mathcal{E}}
\safemath{\setF}{\mathcal{F}}
\safemath{\setG}{\mathcal{G}}
\safemath{\setH}{\mathcal{H}}
\safemath{\setI}{\mathcal{I}}
\safemath{\setJ}{\mathcal{J}}
\safemath{\setK}{\mathcal{K}}
\safemath{\setL}{\mathcal{L}}
\safemath{\setM}{\mathcal{M}}
\safemath{\setN}{\mathcal{N}}
\safemath{\setO}{\mathcal{O}}
\safemath{\setP}{\mathcal{P}}
\safemath{\setQ}{\mathcal{Q}}
\safemath{\setR}{\mathcal{R}}
\safemath{\setS}{\mathcal{S}}
\safemath{\setT}{\mathcal{T}}
\safemath{\setU}{\mathcal{U}}
\safemath{\setV}{\mathcal{V}}
\safemath{\setW}{\mathcal{W}}
\safemath{\setX}{\mathcal{X}}
\safemath{\setY}{\mathcal{Y}}
\safemath{\setZ}{\mathcal{Z}}
\safemath{\emptySet}{\varnothing}

\safemath{\colA}{\mathscr{A}}
\safemath{\colB}{\mathscr{B}}
\safemath{\colC}{\mathscr{C}}
\safemath{\colD}{\mathscr{D}}
\safemath{\colE}{\mathscr{E}}
\safemath{\colF}{\mathscr{F}}
\safemath{\colG}{\mathscr{G}}
\safemath{\colH}{\mathscr{H}}
\safemath{\colI}{\mathscr{I}}
\safemath{\colJ}{\mathscr{J}}
\safemath{\colK}{\mathscr{K}}
\safemath{\colL}{\mathscr{L}}
\safemath{\colM}{\mathscr{M}}
\safemath{\colN}{\mathscr{N}}
\safemath{\colO}{\mathscr{O}}
\safemath{\colP}{\mathscr{P}}
\safemath{\colQ}{\mathscr{Q}}
\safemath{\colR}{\mathscr{R}}
\safemath{\colS}{\mathscr{S}}
\safemath{\colT}{\mathscr{T}}
\safemath{\colU}{\mathscr{U}}
\safemath{\colV}{\mathscr{V}}
\safemath{\colW}{\mathscr{W}}
\safemath{\colX}{\mathscr{X}}
\safemath{\colY}{\mathscr{Y}}
\safemath{\colZ}{\mathscr{Z}}

\safemath{\opA}{\mathbb{A}}
\safemath{\opB}{\mathbb{B}}
\safemath{\opC}{\mathbb{C}}
\safemath{\opD}{\mathbb{D}}
\safemath{\opE}{\mathbb{E}}
\safemath{\opF}{\mathbb{F}}
\safemath{\opG}{\mathbb{G}}
\safemath{\opH}{\mathbb{H}}
\safemath{\opI}{\mathbb{I}}
\safemath{\opJ}{\mathbb{J}}
\safemath{\opK}{\mathbb{K}}
\safemath{\opL}{\mathbb{L}}
\safemath{\opM}{\mathbb{M}}
\safemath{\opN}{\mathbb{N}}
\safemath{\opO}{\mathbb{O}}
\safemath{\opP}{\mathbb{P}}
\safemath{\opQ}{\mathbb{Q}}
\safemath{\opR}{\mathbb{R}}
\safemath{\opS}{\mathbb{S}}
\safemath{\opT}{\mathbb{T}}
\safemath{\opU}{\mathbb{U}}
\safemath{\opV}{\mathbb{V}}
\safemath{\opW}{\mathbb{W}}
\safemath{\opX}{\mathbb{X}}
\safemath{\opY}{\mathbb{Y}}
\safemath{\opZ}{\mathbb{Z}}
\safemath{\opZero}{\mathbb{O}}
\safemath{\identityop}{\opI}

\safemath{\veca}{\bma}
\safemath{\vecb}{\bmb}
\safemath{\vecc}{\bmc}
\safemath{\vecd}{\bmd}
\safemath{\vece}{\bme}
\safemath{\vecf}{\bmf}
\safemath{\vecg}{\bmg}
\safemath{\vech}{\bmh}
\safemath{\veci}{\bmi}
\safemath{\vecj}{\bmj}
\safemath{\veck}{\bmk}
\safemath{\vecl}{\bml}
\safemath{\vecm}{\bmm}
\safemath{\vecn}{\bmn}
\safemath{\veco}{\bmo}
\safemath{\vecp}{\bmp}
\safemath{\vecq}{\bmq}
\safemath{\vecr}{\bmr}
\safemath{\vecs}{\bms}
\safemath{\vect}{\bmt}
\safemath{\vecu}{\bmu}
\safemath{\vecv}{\bmv}
\safemath{\vecw}{\bmw}
\safemath{\vecx}{\bmx}
\safemath{\vecy}{\bmy}
\safemath{\vecz}{\bmz}

\safemath{\veczero}{\bmzero}
\safemath{\vecone}{\bmone}
\safemath{\vecxi}{\bmxi}
\safemath{\veclambda}{\bmlambda}
\safemath{\vecmu}{\bmmu}
\safemath{\vectheta}{\bmtheta}
\safemath{\vecphi}{\bmphi}
\safemath{\vecdelta}{\bmdelta}

\safemath{\matA}{\bA}
\safemath{\matB}{\bB}
\safemath{\matC}{\bC}
\safemath{\matD}{\bD}
\safemath{\matE}{\bE}
\safemath{\matF}{\bF}
\safemath{\matG}{\bG}
\safemath{\matH}{\bH}
\safemath{\matI}{\bI}
\safemath{\matJ}{\bJ}
\safemath{\matK}{\bK}
\safemath{\matL}{\bL}
\safemath{\matM}{\bM}
\safemath{\matN}{\bN}
\safemath{\matO}{\bO}
\safemath{\matP}{\bP}
\safemath{\matQ}{\bQ}
\safemath{\matR}{\bR}
\safemath{\matS}{\bS}
\safemath{\matT}{\bT}
\safemath{\matU}{\bU}
\safemath{\matV}{\bV}
\safemath{\matW}{\bW}
\safemath{\matX}{\bX}
\safemath{\matY}{\bY}
\safemath{\matZ}{\bZ}
\safemath{\matzero}{\bmzero}

\safemath{\matDelta}{\bDelta}
\safemath{\matLambda}{\bLambda}
\safemath{\matPhi}{\bPhi}
\safemath{\matSigma}{\bSigma}
\safemath{\matOmega}{\bOmega}
\safemath{\matTheta}{\bTheta}

\safemath{\matidentity}{\matI}
\safemath{\matone}{\matO}

\safemath{\rnda}{A}
\safemath{\rndb}{B}
\safemath{\rndc}{C}
\safemath{\rndd}{D}
\safemath{\rnde}{E}
\safemath{\rndf}{F}
\safemath{\rndg}{G}
\safemath{\rndh}{H}
\safemath{\rndi}{I}
\safemath{\rndj}{J}
\safemath{\rndk}{K}
\safemath{\rndl}{L}
\safemath{\rndm}{M}
\safemath{\rndn}{N}
\safemath{\rndo}{O}
\safemath{\rndp}{P}
\safemath{\rndq}{Q}
\safemath{\rndr}{R}
\safemath{\rnds}{S}
\safemath{\rndt}{T}
\safemath{\rndu}{U}
\safemath{\rndv}{V}
\safemath{\rndw}{W}
\safemath{\rndx}{X}
\safemath{\rndy}{Y}
\safemath{\rndz}{Z}

\safemath{\rveca}{\bimA}
\safemath{\rvecb}{\bimB}
\safemath{\rvecc}{\bimC}
\safemath{\rvecd}{\bimD}
\safemath{\rvece}{\bimE}
\safemath{\rvecf}{\bimF}
\safemath{\rvecg}{\bimG}
\safemath{\rvech}{\bimH}
\safemath{\rveci}{\bimI}
\safemath{\rvecj}{\bimJ}
\safemath{\rveck}{\bimK}
\safemath{\rvecl}{\bimL}
\safemath{\rvecm}{\bimM}
\safemath{\rvecn}{\bimN}
\safemath{\rveco}{\bomO}
\safemath{\rvecp}{\bimP}
\safemath{\rvecq}{\bimQ}
\safemath{\rvecr}{\bimR}
\safemath{\rvecs}{\bimS}
\safemath{\rvect}{\bimT}
\safemath{\rvecu}{\bimU}
\safemath{\rvecv}{\bimV}
\safemath{\rvecw}{\bimW}
\safemath{\rvecx}{\bimX}
\safemath{\rvecy}{\bimY}
\safemath{\rvecz}{\bimZ}

\safemath{\rvecxi}{\bmxi}
\safemath{\rveclambda}{\bmlambda}
\safemath{\rvecmu}{\bmmu}
\safemath{\rvectheta}{\bmtheta}
\safemath{\rvecphi}{\bmphi}

\safemath{\rmatA}{\bimA}
\safemath{\rmatB}{\bimB}
\safemath{\rmatC}{\bimC}
\safemath{\rmatD}{\bimD}
\safemath{\rmatE}{\bimE}
\safemath{\rmatF}{\bimF}
\safemath{\rmatG}{\bimG}
\safemath{\rmatH}{\bimH}
\safemath{\rmatI}{\bimI}
\safemath{\rmatJ}{\bimJ}
\safemath{\rmatK}{\bimK}
\safemath{\rmatL}{\bimL}
\safemath{\rmatM}{\bimM}
\safemath{\rmatN}{\bimN}
\safemath{\rmatO}{\bimO}
\safemath{\rmatP}{\bimP}
\safemath{\rmatQ}{\bimQ}
\safemath{\rmatR}{\bimR}
\safemath{\rmatS}{\bimS}
\safemath{\rmatT}{\bimT}
\safemath{\rmatU}{\bimU}
\safemath{\rmatV}{\bimV}
\safemath{\rmatW}{\bimW}
\safemath{\rmatX}{\bimX}
\safemath{\rmatY}{\bimY}
\safemath{\rmatZ}{\bimZ}

\safemath{\rmatDelta}{\bimDelta}
\safemath{\rmatLambda}{\bimLambda}
\safemath{\rmatPhi}{\bimPhi}
\safemath{\rmatSigma}{\bimSigma}
\safemath{\rmatOmega}{\bimOmega}
\safemath{\rmatTheta}{\bimTheta}

\usepackage{amssymb}
\usepackage{amsfonts}
\usepackage{mathrsfs}
\usepackage{xspace}
\usepackage{bm}
\usepackage{fancyref}
\usepackage{textcomp}

\usepackage{multirow}
\usepackage{stmaryrd}

\newenvironment{textbmatrix}{	\setlength{\arraycolsep}{2.5pt}%
								\big[\begin{matrix}}{\end{matrix}\big]%
								\raisebox{0.08ex}{\vphantom{M}}}

\def\be{\begin{equation}}
\def\ee{\end{equation}}
\def\een{\nonumber \end{equation}}
\def\mat{\begin{bmatrix}}
\def\emat{\end{bmatrix}}
\def\btm{\begin{textbmatrix}}
\def\etm{\end{textbmatrix}}

\def\ba#1\ea{\begin{align}#1\end{align}}
\def\bas#1\eas{\begin{align*}#1\end{align*}}
\def\bs#1\es{\begin{split}#1\end{split}} 
\def\bg#1\eg{\begin{gather}#1\end{gather}}
\def\bml#1\eml{\begin{multline}#1\end{multline}}
\def\bi#1\ei{\begin{itemize}#1\end{itemize}}

\safemath{\dirac}{\delta}					
\safemath{\krond}{\dirac}					

\safemath{\upto}{\uparrow}
\safemath{\downto}{\downarrow}
\safemath{\iu}{j}							
\safemath{\ev}{\lambda}						
\safemath{\hilseqspace}{l^{2}}				
\newcommand{\banachfunspace}[1]{\setL^{#1}}	
\safemath{\hilfunspace}{\banachfunspace{2}}	

\safemath{\SNR}{\text{\sc snr}} 				
\safemath{\No}{N_0}							
\safemath{\Es}{E_s}							
\safemath{\Eb}{E_b}							
\safemath{\EbNo}{\frac{\Eb}{\No}}
\safemath{\EsNo}{\frac{\Es}{\No}}

\DeclareMathOperator{\CHop}{\ensuremath{\opH}} 
\safemath{\tvir}{\rndh_{\CHop}}				
\safemath{\tvtf}{\rndl_{\CHop}}				
\safemath{\spf}{\rnds_{\CHop}}				
\safemath{\bff}{H_{\CHop}}					

\safemath{\ircf}{r_{h}}						
\safemath{\tftvcf}{r_{s}}					
\safemath{\tfcf}{r_{l}}						
\safemath{\bfcf}{r_{H}}						

\safemath{\tcorr}{c_h}						
\safemath{\scf}{c_{s}}						
\safemath{\tfcorr}{c_{l}}					
\safemath{\fcorr}{c_{H}}						

\safemath{\mi}{I}							
\safemath{\capacity}{C}						

\safemath{\normal}{\mathcal{N}}			
\safemath{\jpg}{\mathcal{CN}}			
\safemath{\mchain}{\leftrightarrow}		

\safemath{\dB}{\,\mathrm{dB}}
\safemath{\dBm}{\,\mathrm{dBm}}
\safemath{\Hz}{\,\mathrm{Hz}}
\safemath{\kHz}{\,\mathrm{kHz}}
\safemath{\MHz}{\,\mathrm{MHz}}
\safemath{\GHz}{\,\mathrm{GHz}}
\safemath{\s}{\,\mathrm{s}}
\safemath{\ms}{\,\mathrm{ms}}
\safemath{\mus}{\,\mathrm{\text{\textmu}s}}
\safemath{\ns}{\,\mathrm{ns}}
\safemath{\ps}{\,\mathrm{ps}}
\safemath{\meter}{\,\mathrm{m}}
\safemath{\mm}{\,\mathrm{mm}}
\safemath{\cm}{\,\mathrm{cm}}
\safemath{\W}{\,\mathrm{W}}
\safemath{\mW}{\, \mathrm{mW}}
\safemath{\J}{\,\mathrm{J}}
\safemath{\K}{\,\mathrm{K}}
\safemath{\bit}{\,\mathrm{bit}}
\safemath{\nat}{\,\mathrm{nat}}

\safemath{\define}{\triangleq}			

\safemath{\equivalent}{\sim}
\safemath{\distas}{\sim}					
\safemath{\sdiff}{\Delta}				

\safemath{\reals}{\mathbb{R}}
\safemath{\positivereals}{\reals_{+}}
\safemath{\integers}{\mathbb{Z}}
\safemath{\posint}{\integers_{+}}
\safemath{\naturals}{\mathbb{N}}
\safemath{\posnaturals}{\naturals_{+}}
\safemath{\complexset}{\mathbb{C}}
\safemath{\rationals}{\mathbb{Q}}

\newcommand*{\fancyrefapplabelprefix}{app}		
\newcommand*{\fancyrefthmlabelprefix}{thm}		
\newcommand*{\fancyreflemlabelprefix}{lem}		
\newcommand*{\fancyrefcorlabelprefix}{cor}		
\newcommand*{\fancyrefdeflabelprefix}{def}		
\newcommand*{\fancyrefproplabelprefix}{prop}		
\newcommand*{\fancyrefexmpllabelprefix}{exmpl}
\newcommand*{\fancyreftbllabelprefix}{tbl}		
\frefformat{vario}{\fancyrefseclabelprefix}{Section~#1}
\frefformat{vario}{\fancyrefthmlabelprefix}{Theorem~#1}
\frefformat{vario}{\fancyreflemlabelprefix}{Lemma~#1}
\frefformat{vario}{\fancyrefcorlabelprefix}{Corollary~#1}
\frefformat{vario}{\fancyrefdeflabelprefix}{Definition~#1}
\frefformat{vario}{\fancyreffiglabelprefix}{Fig.~#1}
\frefformat{vario}{\fancyreftbllabelprefix}{Tbl.~#1}
\frefformat{vario}{\fancyrefapplabelprefix}{Appendix~#1}
\frefformat{vario}{\fancyrefeqlabelprefix}{(#1)}
\frefformat{vario}{\fancyrefproplabelprefix}{Property~#1}
\frefformat{vario}{\fancyrefexmpllabelprefix}{Example~#1}

\safemath{\ysig}{\bmy}
\safemath{\ysighat}{\hat{\ysig}}
\safemath{\ysigdim}{M}
\safemath{\xsig}{\bmx}
\safemath{\xsigdim}{N}
\safemath{\nx}{n_x}
\safemath{\zsig}{\bmz}
\safemath{\zsigdim}{\ysigdim}
\safemath{\rsig}{\bmr}
\safemath{\Adict}{\bA}
\safemath{\Adicttilde}{\widetilde{\Adict}}
\safemath{\Adictdim}{\outputdim\times\xsigdim}
\safemath{\avec}{\bma}
\safemath{\avectilde}{\tilde{\avec}}
\safemath{\Bdict}{\bB}
\safemath{\Bdicttilde}{\widetilde{\Bdict}}
\safemath{\Cdict}{\bC}
\safemath{\cvec}{\bmc}
\safemath{\Ddict}{\bD}
\safemath{\Ddictdim}{\ysigdim\times\xsigdim}
\safemath{\dvec}{\bmd}
\safemath{\Ddicttilde}{\widetilde{\bD}}
\safemath{\Bonb}{\bB}
\safemath{\bvec}{\bmb}
\safemath{\Bonbdim}{\ysigdim\times\ysigdim}
\safemath{\noise}{\bmn}
\safemath{\noisedim}{\ysigim}
\safemath{\err}{\bme}
\safemath{\errdim}{\ysigdim}
\safemath{\errset}{\setE}
\safemath{\nerr}{n_e}
\safemath{\delop}{\bP_\errset}
\safemath{\delopc}{\bP_{{\errset}^c}}

\safemath{\cplxi}{\imath}
\safemath{\cplxj}{\jmath}

\safemath{\dict}{\matD}
\safemath{\inputdim}{N}		
\safemath{\outputdim}{M}		
\safemath{\sparsity}{s}	
\safemath{\inputdimA}{{N_a}}	
\safemath{\inputdimB}{{N_b}}	
\safemath{\elemA}{{n_a}}	
\safemath{\elemB}{{n_b}}	
\safemath{\resA}{\matR_a}	
\safemath{\resB}{\matR_b}	
\safemath{\subD}{\matS} 
\safemath{\subA}{\matS_a} 
\safemath{\subB}{\matS_b} 
\safemath{\dicta}{\matA} 	
\safemath{\dictb}{\matB} 	
\safemath{\hollowS}{H}
\safemath{\hollowA}{H_a}
\safemath{\hollowB}{H_b}
\safemath{\cross}{Z}
\safemath{\coh}{\mu}			
\safemath{\coha}{\mu_a}			
\safemath{\cohb}{\mu_b}			
\safemath{\mubs}{\nu}	
\safemath{\cohm}{\mu_m} 
\safemath{\dictset}{\setD}	
\safemath{\dictsetp}{\dictset(\coh,\coha,\cohb)}	
\safemath{\dictsetgen}{\dictset_\text{gen}}
\safemath{\dictsetgenp}{\dictsetgen(\coh)}
\safemath{\dictsetonb}{\dictset_\text{onb}}
\safemath{\dictsetonbp}{\dictsetonb(\coh)}

\safemath{\leftside}{U}
\safemath{\rightsideA}{R_a}
\safemath{\rightsideB}{R_b}

\safemath{\indexS}{\setI_S} 

\safemath{\na}{n_a}			
\safemath{\nb}{n_b}			
\safemath{\coeffa}{p_i}	
\safemath{\coeffb}{q_j}	
\safemath{\seta}{\setP}		
\safemath{\setb}{\setQ}     
\safemath{\setw}{\setW}	
\safemath{\setz}{\setZ}	
\safemath{\cola}{\veca}		
\safemath{\colb}{\vecb}		
\safemath{\cold}{\vecd}		
\safemath{\inputvec}{\vecx} 	
\safemath{\error}{\vece}	
\safemath{\noiseout}{\vecz} 	
\safemath{\inputvecel}{x}
\safemath{\inputveca}{\vecx_a}
\safemath{\inputvecb}{\vecx_b}
\safemath{\outputvec}{\vecy}	
\safemath{\lambdamin}{\lambda_{\mathrm{min}}}

\safemath{\elltwo}{\ell_2}
\safemath{\ellone}{\ell_1}
\safemath{\ellzero}{\ell_0}
\safemath{\ellinf}{\ell_\infty}
\safemath{\licard}{Z(\coh,\coha,\cohb)}
\safemath{\xsol}{\hat{x}}
\safemath{\xbord}{x_b}		
\safemath{\xstat}{x_s}		
\safemath{\xstatLone}{\tilde{x}_s}
\safemath{\order}{\mathcal{O}} 
\safemath{\scales}{\Theta} 
\safemath{\ones}{\mathbf{1}} 
\safemath{\zeroes}{\mathbf{0}} 
\safemath{\thlone}{\kappa(\coh,\cohb)} 
\safemath{\constoneA}{\delta} 
\safemath{\constoneB}{\epsilon} 
\safemath{\nlarge}{L}				   
\safemath{\sumlarge}{S_\nlarge}
\safemath{\maxlarger}{P_\nlarge}	   
\safemath{\Pzero}{\textrm{P0}}	
\safemath{\Pone}{\textrm{P1}}
\safemath{\vecfir}{\vecw}			 
\safemath{\vecsec}{\vecz}
\safemath{\elvecfir}{w}              
\safemath{\elvecsec}{z}				 
\safemath{\nlargefir}{n}
\safemath{\normout}{\gamma}
\safemath{\auxfun}{h}
\safemath{\supp}{\textrm{supp}}

\safemath{\indexa}{\ell}
\safemath{\indexb}{r}
\safemath{\indexc}{i}
\safemath{\indexd}{j}

\safemath{\project}{P}

\usepackage{color}
\usepackage{enumitem}
\setlist[itemize]{leftmargin=*, itemsep=0.3em, topsep=0.3em} 
\allowdisplaybreaks 
\usepackage{amssymb}
\usepackage{tabularx}
\title{Supervised Contrastive CSI Representation \\ Learning for Massive MIMO Positioning}
\author{Junquan Deng, Wei Shi, Jianzhao Zhang,  Xianyu Zhang, and Chuan Zhang
\thanks{JD, WS, JZ and XZ are with Sixty-third Research Institute, National University of Defense Technology, Nanjing, China~(e-mail: jqdeng@nudt.edu.cn; w.shi@nudt.edu.cn; jianzhao63s@nudt.edu.cn; zhangxy\_sat@126.com). CZ is with LEADS of Southeast University, and Purple Mountain Laboratories, Nanjing, China~(e-mail: chzhang@seu.edu.cn).
This work is supported by NSFC under grant 61901497 and 62131005, in part by China Postdoctoral Science Foundation~(No. 2021MD703980) and Research Project of NUDT under grant ZK 19-09.}
}
\begin{document}

\maketitle

\begin{abstract}
%
Similarity metric is crucial for massive MIMO positioning utilizing channel state information~(CSI).
In this letter, we propose a novel massive MIMO CSI similarity learning method via deep convolutional neural network~(DCNN) and contrastive learning. A contrastive loss function is designed considering multiple positive and negative CSI samples drawn from a training dataset. The DCNN encoder is trained using the loss so that positive samples are mapped to points close to the anchor's encoding, while encodings of negative samples are kept away from the anchor's in the representation space. Evaluation results of fingerprint-based positioning on a real-world CSI dataset show that the learned similarity metric improves positioning accuracy significantly compared with other known state-of-the-art methods.
\end{abstract}


\section{Introduction}
\label{sec:intro}
\IEEEPARstart{A}{quiring} location information of mobile devices is essential in many smart city and internet-of-things~(IoT) applications, including traffic
monitoring, asset tracking, autonomous driving, emergency rescue and so on.
%
%
5G and beyond wireless networks feature the widely use of massive multiple-input multiple output~(MIMO) transmission technique, which not only increases the spectrum efficiencies of wireless transmissions, but also equips the networks with higher sensing and positioning capabilities.
Massive MIMO channel state information~(CSI) with channel responses on multiple antennas, has been utilized for positioning using geometrical methods~\cite{NSDI13_ArrayTrack,Wymeersch_2017,TWC2018_Mendrzik}, fingerprinting~\cite{CSIMIMO_ICC2020,JCC2021,Luc_LCOMM2021,TVT2019_sun,Wu_TWC_2021,PIMRC2017_Tufvesson,Li2021TVT,Studer_TWC2021}, channel charting~(CC)~\cite{CC2018,MPCC2018,Siamese2019,IWCMC2021,ICCC2021,Ferrand2021} and direct mapping via deep neural network~(DNN)~\cite{GLOBECOM2020_DNN,Bast_VTC_2020,EuCNC2021}. Geometrical methods are based on Direction-of-Arrival~(DOA) and  Time-of-Arrival~(TOA) estimates, which require rigorous array calibration and accurate synchronization among distributed network entities.  Fingerprinting, CC and direct mapping, alleviate these challenges, by using a labeled dataset and  machine learning methods, including k-nearest neighbors~(kNN), supervised dimensionality reduction and deep neural network, to predict CSI's associated locations. CSI similarity metric plays a vital role in fingerprinting and CC-based positioning methods. In fact, the massive MIMO positioning methods in~\cite{CSIMIMO_ICC2020,JCC2021,Luc_LCOMM2021,TVT2019_sun,Wu_TWC_2021,PIMRC2017_Tufvesson,Li2021TVT,Studer_TWC2021,CC2018,Siamese2019,IWCMC2021,ICCC2021,Ferrand2021} build on the assumption that two CSI samples measured at nearby locations should be close to each other with a specific similarity metric.

A joint angle-delay similarity coefficient was proposed in~\cite{TVT2019_sun}, and correlation matrix distance~(CMD) was applied in~\cite{Wu_TWC_2021}, both using the angle-delay channel power matrix (ADCPM) as CSI feature. In~\cite{IWCMC2021,ICCC2021}, the CSI similarity metric is based on power angular profile (PAP), which is estimated by a multiple signal classification~(MUSIC) algorithm. In~\cite{CC2018}, the Frobenius distance between channel covariance matrices scaled by a vary factor is adopted to measure CSI difference. The above-mentioned CSI similarity metrics assume antenna arrays with perfect linear structures and calibration, making them questionable for practical massive MIMO systems. Other hand-crafted CSI similarity metrics have been designed for general MIMO systems. For example, average Euclidean distance between CSI magnitude values over multiple antennas and sub-carriers is used in~\cite{CSIMIMO_ICC2020,Sobehy2021}. Furthermore, the left singular vector corresponding to the largest singular value of MIMO channel frequency response matrix is selected as the CSI feature in~\cite{Luc_LCOMM2021}, and inner product of such features is used to measure the CSI similarity; Such a similarity metric is closely related to Chordal distance~\cite{Chordal_2012} on the Grassmannian manifold.

The CSI estimated at the MIMO receiver from a transmitter position is affected by the surrounding environment and the characteristics of radio frequency~(RF) chains, which are hard to be parameterized by a deterministic model. Direct mapping via DNN tries to learn a function $\mathbf{\bar{p}}= \textrm{Proj}(\rm{CSI})$ to predict the position $\mathbf{p}$ from CSI, which inherently describes the uncertain environment and RF characteristics. However, to train reliable DNNs, dense CSI samples are needed~\cite{GLOBECOM2020_DNN,Bast_VTC_2020,EuCNC2021}, and their positioning accuracies are inferior to classical kNN methods as indicated in~\cite{Rashdan2020,GarauBurguera2020,Sobehy2021}, especially when the training CSI samples are sparse.

In this letter, we propose a new  CSI similarity model based on contrastive learning~\cite{chen2020simple,NEURIPS2020}, for single-site massive MIMO positioning. The goal is to learn reliable CSI features so that CSI samples collected at neighboring locations are projected by a deep convolutional neural network~(DCNN) to neighboring points in a intermediate Euclidean feature space. In the leaning phase, we construct positive and negative CSI samples for each CSI based on their location information in the training dataset. The DCNN is trained using a novel contrastive loss, which considers multiple positive and  multiple negative samples. The advantages of our proposed CSI similarity model are:  \textbf{i)} universally applicable to different wireless systems with different types of CSI data as long as  spatial consistency~\cite{WCNC2019_Kurras} holds for the wireless channels; \textbf{ii)} no specific knowledge of the antenna array structure and array calibration are needed; \textbf{iii)} adaptive to complex radio environment with non-line-of-sight~(NLOS) and multi-path conditions. The code is released at \emph{https://github.com/dengjunquan/SupConCSI}.

\section{System Model and Problem Formulation}
\label{sec:systemmodel}
\begin{figure}[tp]
\centering
\includegraphics[width=0.48\textwidth]{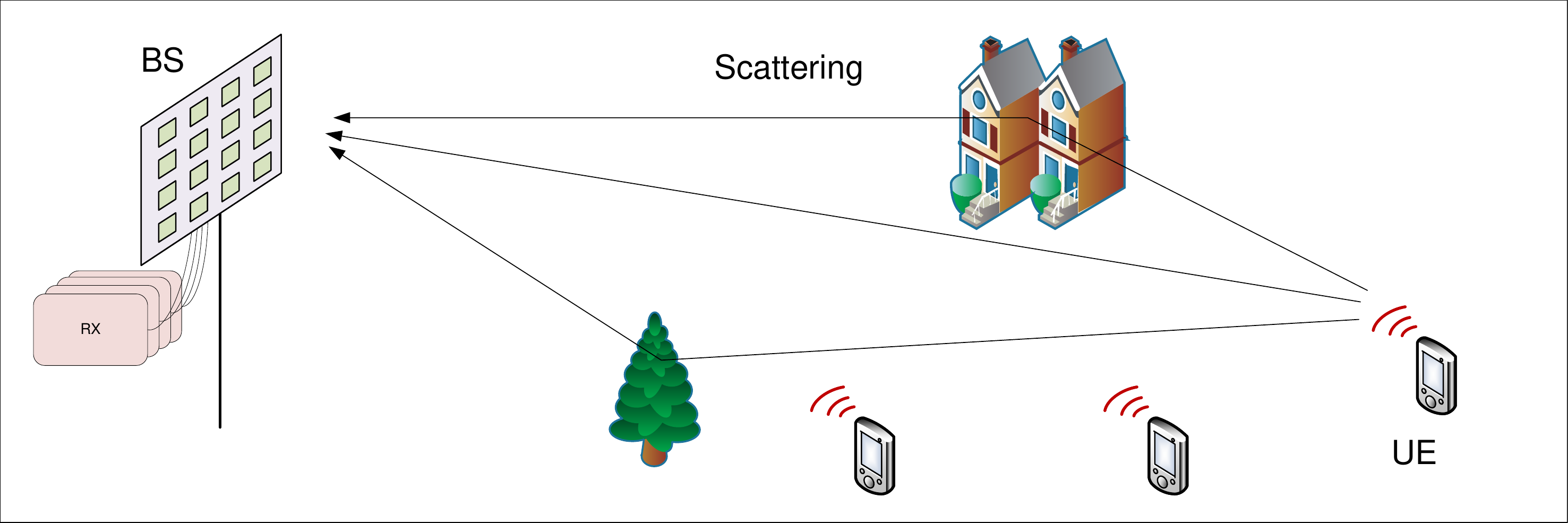}
\caption{Single-site massive MIMO positioning scenario.}
\label{fig:Scenario}
\end{figure}

We consider a single-site massive MIMO positioning system, where a base station~(BS) with $B$ antennas receives pilot signals from single-antenna user equipments~(UEs), as depicted in Fig.~\ref{fig:Scenario}. A typical MIMO orthogonal frequency division multiplexing~(OFDM) air interface is adopted, and the raw CSI from an UE is represented by the estimated spatial-frequency channel response matrix as
\begin{equation}\label{eq:CSI}
\begin{aligned}
\mb H = \mb H_o+ \mb H_{e} = [\mb h_0, \mb h_1,\cdots,\mb h_{N-1} ] + \mb H_{e} \in \bb C^{B\times N},
\end{aligned}
\end{equation}
where $N$ is the number of OFDM sub-carriers, $\mb h_i \in \bb C^{B\times 1}$ is the channel response vector on the $i$th sub-carrier, and $\mb H_{e}$ represents the channel estimation error. The values of $\mb H$ are complex number and affected by a range of factors, including powers, DOAs and delays of multi-path components, and antenna array structure, antenna pattern \& coupling effects, carrier frequency offset (CFO), symbol timing offset (STO)~\cite{Gauger2020}, which are difficult to be parameterized in practical systems. We make no assumption on details of these factors, and assume that the relationship between $\mb H$ and the UE position $\mb p$ can be viewed as a black box model $\mb H = \mc G (\mb p,\mc X)$, with $\mc X$ representing the unknown factors.

In a fingerprint-based massive MIMO positioning system, a database consist of CSI samples and their corresponding UE positions is first constructed via a dedicated site survey or crowdsourcing. Denoting the database as $\{ \mb H_i, \mb p_i\}_{i=1\ldots I}$, our goal is to learn a similarity metric ${\rm S}(\mb H_i, \mb H_j)$  for $i \neq j$, such that if $\mb p_j$ is a k-nearest neighbor of $\mb p_i$ ($i = 1\ldots I$) in the geographical space, $\mb H_j$ should also be a k-nearest neighbor of $\mb H_i$ with this metric. This should hold for a new CSI measurement, i.e., if $\mb H$ is measured from an unknown location $\mb p$ and $\mb p_j$~($j =1\ldots I$) is a k-nearest neighbor of $\mb p$, $\mb H_j$ should also be a k-nearest neighbor of $\mb H$.

\section{Contrastive CSI Representation Learning}

Contrastive learning~(CL) methods have achieved great successes in computer vision~\cite{chen2020simple,NEURIPS2020}, its goal is to learn a representing space where positive samples stay close to the anchor sample, while negative ones are far apart. The mapping function is typically implemented by a neural network.  For each anchor, its positives can be generated using data augmentation techniques in an unsupervised setting~\cite{chen2020simple} or leveraging the available labels in a supervised version~\cite{NEURIPS2020}, while its negatives can be drawn randomly from the sample set.

\begin{figure}[tp]
\centering
\includegraphics[width=0.48\textwidth]{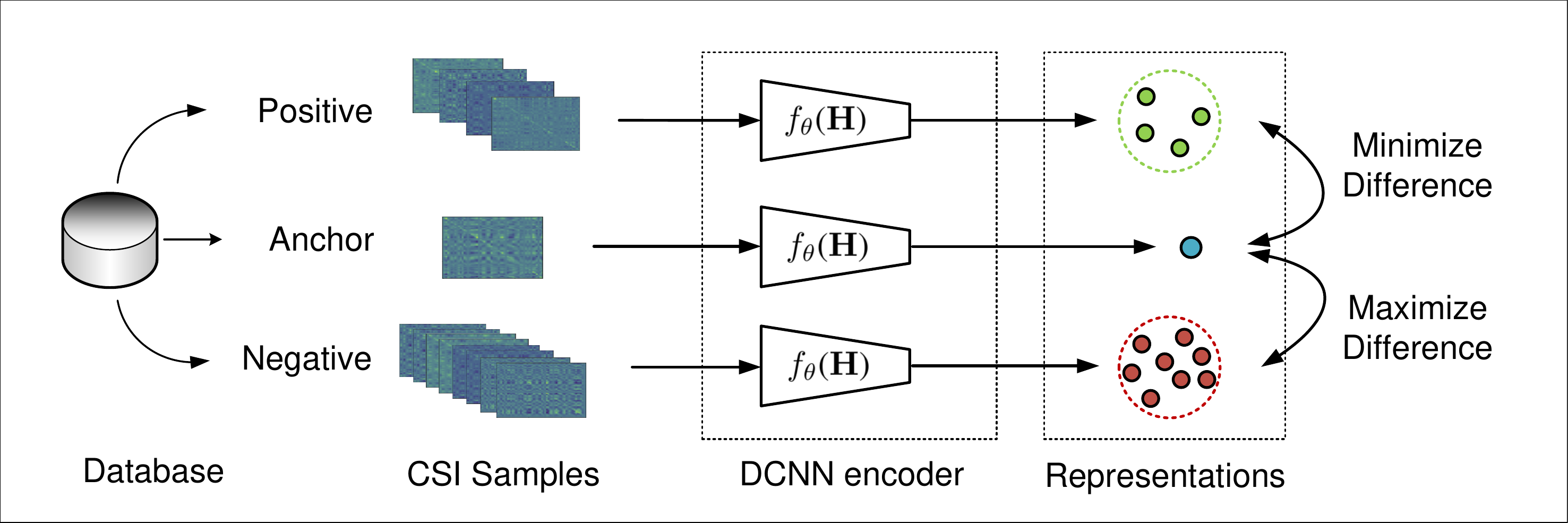}
\caption{Contrastive CSI  representation learning pipeline.}
\label{fig:ConSCI}
\end{figure}

Following a typical CL procedure as in~\cite{NEURIPS2020}, we now detail the contrastive CSI  representation learning pipeline for massive MIMO positioning, as shown in Fig.~\ref{fig:ConSCI}. It has three key steps: \textbf{i)} construction of positive and negative samples based on the available position information for an anchor CSI; \textbf{ii)} map the CSI samples to the feature space via an encoder $\mb z = f_{\theta}$($\mb H$); \textbf{iii)} use a contrastive loss and an optimizer to update the weights of the encoder.

\subsection{Construction of Positive and Negative Samples}
For a CSI sample $\mb H_{a}$ in a mini-batch  $\{ \mb H_{a}, \mb p_{a}\}_{a\in \mc A}$ of the training dataset $\{ \mb H_i, \mb p_i\}_{i=1\ldots I}$, its positive samples $\{ \mb H_{p}\}_{p\in \mc P_a}$ are the $|\mc P_a|$ nearest samples in the training dataset according to Euclidean distance $||\mb p_i - \mb p_{a}||_2$. Meanwhile, the negative ones $\{ \mb H_{n}\}_{n \in \mc N_a}$  are randomly selected from the training dataset with $||\mb p_i - \mb p_{a}||_2 > d_{\rm {th}}$, with $d_{\rm {th}}$ is a predefined distance threshold. In what follows, the mini-batch, positive and negative samples are represented by the index sets $\mc A$, $\mc P_a$, $\mc N_a \subset \{1 \ldots I\}$.
\subsection{DCNN-Based Encoder Network}
\begin{figure}[hp]
\centering
\includegraphics[width=0.48\textwidth]{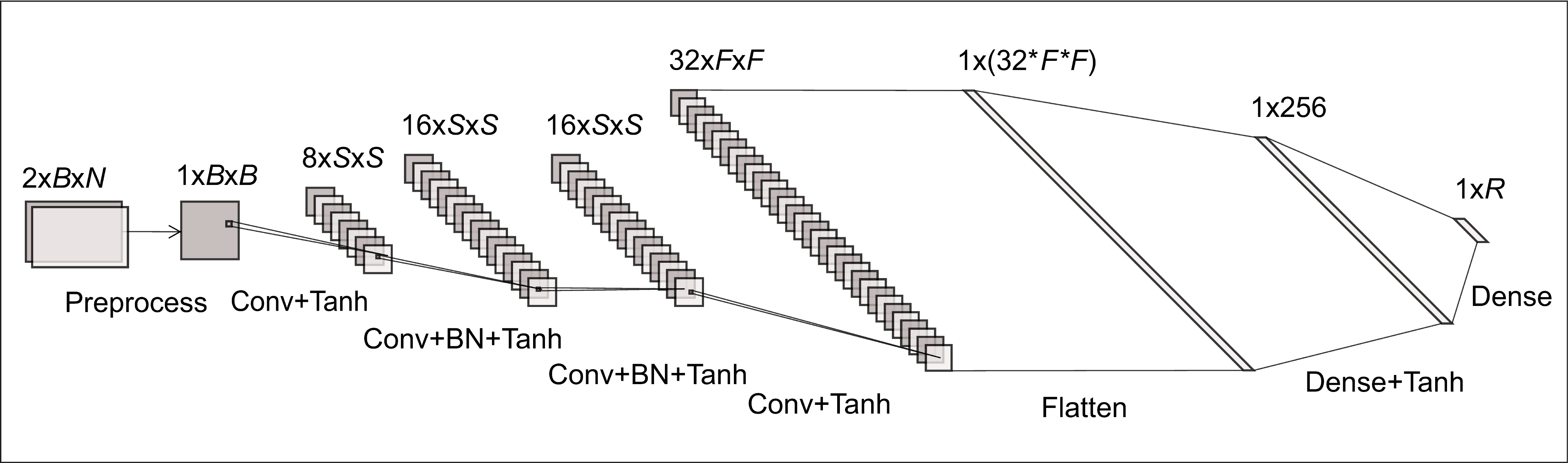}
\caption{Proposed DCNN-based CSI encoder $f_{\theta}$($\mb H$).}
\label{fig:DCNN}
\end{figure}

In order to learn robust intermediate CSI features for positioning purpose, we have designed a DCNN-based encoder network $f_{\theta}$($\mb H$) as depicted in Fig.~\ref{fig:DCNN}, which is consist of a fixed CSI preprocessing layer, a DCNN block with 2D convolution layers and a projection head with two dense layers. The dimension of output feature is $1\times R$ and is much smaller than the original CSI size of $2\times B\times N$ as the input $\mb H$ is a complex matrix with real and imaginary parts.
Noticing that autocorrelation over spatial or delay domains are more stable then the raw MIMO-OFDM channel response matrix~\cite{IWCMC2021,Studer_TWC2021}, in the preprocessing layer we first compute the autocorrelation matrix $\mb C = \mb H \mb H^{\dagger}$. As $\mb C$ is Hermitian, we then use
\begin{equation}\label{eq:R}
\begin{aligned}
\mb R = \mathfrak{Re}\{\rm{triu}(\mb C)\} + \sqrt{-1}\, \mathfrak{Im}\{ \rm{tril}(\mb C)\} + \rm{diag}(\mb C)
\end{aligned}
\end{equation}
as the input for convolution layers, with $\rm{triu}(\cdot),\rm{tril}(\cdot ),\rm{diag}(\cdot)$ denoting upper triangular, lower triangular and diagonal parts of a matrix, and $\mathfrak{Re}\{\cdot\}$, $\mathfrak{Im}\{\cdot\}$ the real and imaginary parts of a matrix.

The DCNN block has four 2D convolution layers, each followed by an activation layer. Different from conventional image input for a CNN, here entities of $\mb R$ can be positive or negative, so we choose the \textbf{Tanh} function for all activation layers. To avoid the problems of over-fitting and vanishing or exploding gradient, we add two batch normalization~(BN) layers in the middle as shown in Fig.~\ref{fig:DCNN}. More details of the network parameters are given in Section~\ref{sec:results}.
\subsection{Supervised Contrastive Loss}
Given the CSI encoder network $\mb z = f_\theta(\mb H)$ and the positive and negative samples $\mc P_a$ and $\mc N_a$ related to an anchor $\mb H_{a}$ in a mini-batch $\mc A$ of size $A$, we propose a contrastive loss function which is given by
\begin{equation}\label{eq:Loss}
\resizebox{0.91\hsize}{!}{$
\begin{aligned}
\mathcal{L} = \frac{1}{A} \sum_{a \in \mc A} \left[-\sum_{p\in \mathcal{P}_a} \log \frac{{{e}}^{\mathbf{z}_a \bigcdot\, \mathbf{z}_p/\tau} }{ \sum\limits_{n \in \mathcal{N}_a } {{e}}^{\mathbf{z}_a \bigcdot\, \mathbf{z}_n/\tau} + \sum\limits_{ r \in \mathcal{P}_a} {{e}}^{\mathbf{z}_a \bigcdot\, \mathbf{z}_{r}/\tau}}\right],
\end{aligned}
$}
\end{equation}
where $\mb z_a,\,\mb z_p$ and $\mb z_n$ are the representations of an anchor, positive and negative CSI samples, and the dot symbol denotes inner product.
The fraction before taking logarithm represents the probability that $\mathbf{z}_a$ selects $\mathbf{z}_p$ over all positives and negatives as one of its neighbours in  feature space. In this regard, minimizing $\mathcal{L}$ would make $\mathbf{z}_a$ and $\mathbf{z}_p$ close.
The temperature $\tau$ is used for tuning how concentrated the features are in the feature space. With a low temperature, $\mathcal{L}$ is dominated by the small distances and widely separated features contribute less.

The proposed loss is inspired by both the InfoNCE loss~\cite{infoNCE2018} and the soft nearest neighbor~(SNN) loss~\cite{pmlr-v97-frosst19a}, commonly used in self-supervised representation learning. The differences are that: 1) InfoNCE is suitable for instance discrimination tasks and involves only a single positive sample; 2) In SNN loss, the summation in the numerator over positives is located inside the logarithm operation; Moreover, we use inner product $\mb z_i \bigcdot\, \mb z_j$ instead of $-||\mb z_i -\mb z_j||_2^2$ in SNN~\cite{pmlr-v97-frosst19a} to measure the similarity of the embedded features, which is more stable and efficient for gradient-descent computation.

To give more details of $\mathcal{L}$, we derive its gradient with respect to $\mathbf{z}_a$. We focus on the loss $\mc L_a$ related to $\mathbf{z}_a$, which is expressed as inside the brackets in~\eqref{eq:Loss}. The gradient is given by
\begin{equation}\label{eq:grad}
\begin{aligned}
\frac{\partial \mc L_a}{\partial \mb z_a } = \frac{|\mathcal{P}_a|}{\tau} \left[ \sum_{p\in \mathcal{P}_a} - \left[\frac{1}{|\mathcal{P}_a|} - x_{ap}\right] \mb z_p + \sum_{ n \in \mathcal{N}_a} x_{an} \mb  z_n  \right],
\end{aligned}
\end{equation}
where $x_{ap} ={e^{\mb z_a \bigcdot\, \mb z_p}}\big/\, {\sum_{j \in \mathcal{N}_a \bigcup \mathcal{P}_a } {{e}}^{\mathbf{z}_a \bigcdot\, \mathbf{z}_j/\tau}}$ and the same form for $ x_{an}$. In the initialization phase, the encoder $f_\theta(\mb H)$ would generate random features, then $x_{ap}$ and $x_{an}$ would have values around ${1}/{(|\mathcal{P}_a|+|\mathcal{N}_a|)}$. When there are enough negatives, $x_{ap}$ is smaller than ${1}/{|\mathcal{P}_a|}$ for most of the positives during training, so $\mb z_a$ will be dragged towards the mean positive representation vector, while be pulled away from the negatives via gradient-descent.
With a well-trained encoder $f_{\theta}(\mb H)$, the supervised contrastive learning-based CSI similarity metric~(\textbf{SupCon}) is
\begin{align}\label{eq:supcon}
{\rm S}(\mb H_i, \mb H_j) \stackrel{\text{def}}{=} ||f_{\theta}(\mb H_i) - f_{\theta}(\mb H_j)||_2^{-1} = ||\mb z_i - \mb z_j||_2^{-1},
\end{align}
and will be used for downstream kNN-based positioning tasks.
\section{Experiments}\label{sec:results}
\begin{figure}[tp]
    \centering
    \begin{subfigure}{0.22\textwidth}
		\centering
		\includegraphics[width=\textwidth]{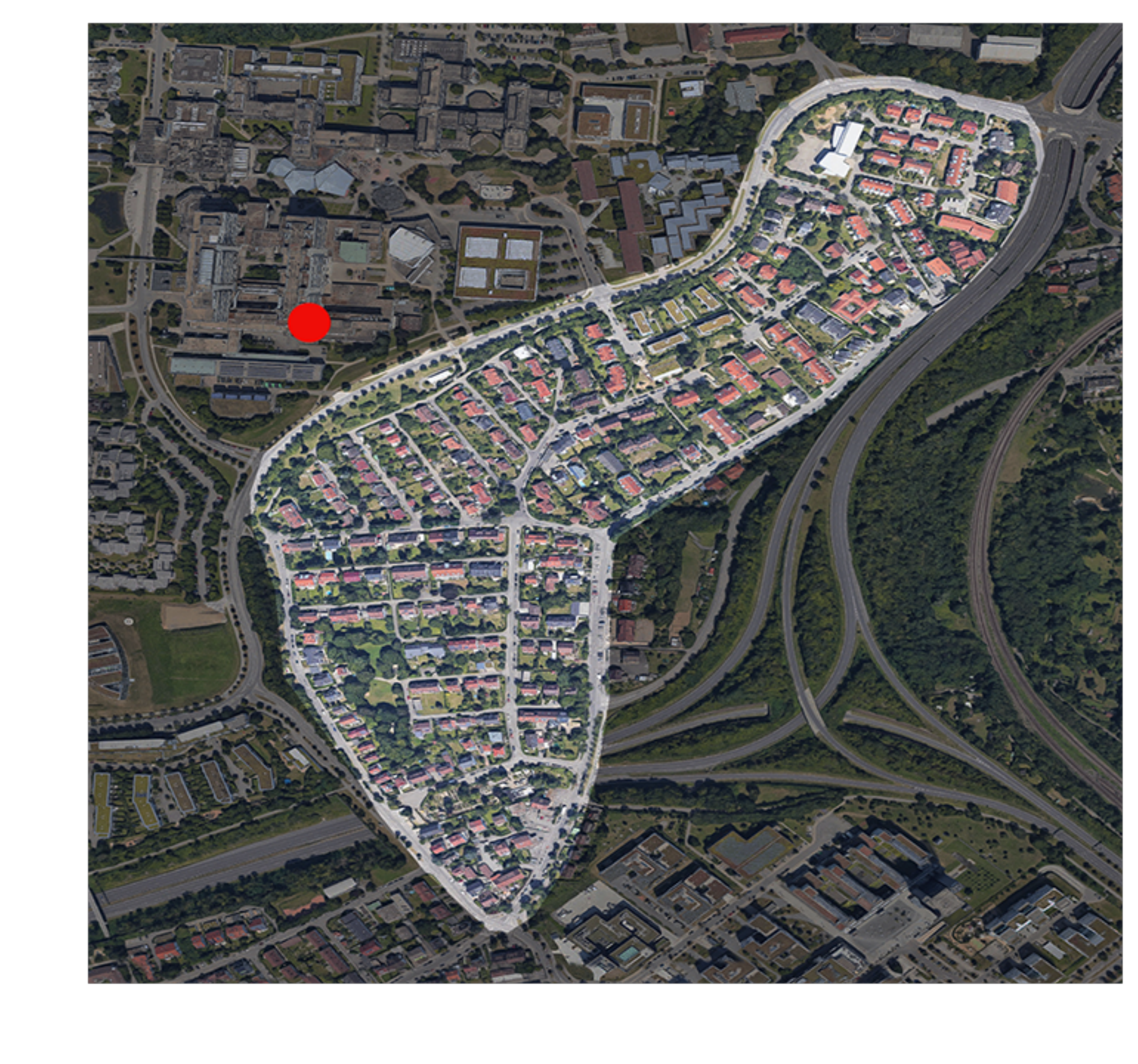}
        \vspace*{-6mm}
		\caption{}
	\end{subfigure}
    \begin{subfigure}{0.22\textwidth}
		\centering
		\includegraphics[width=\textwidth]{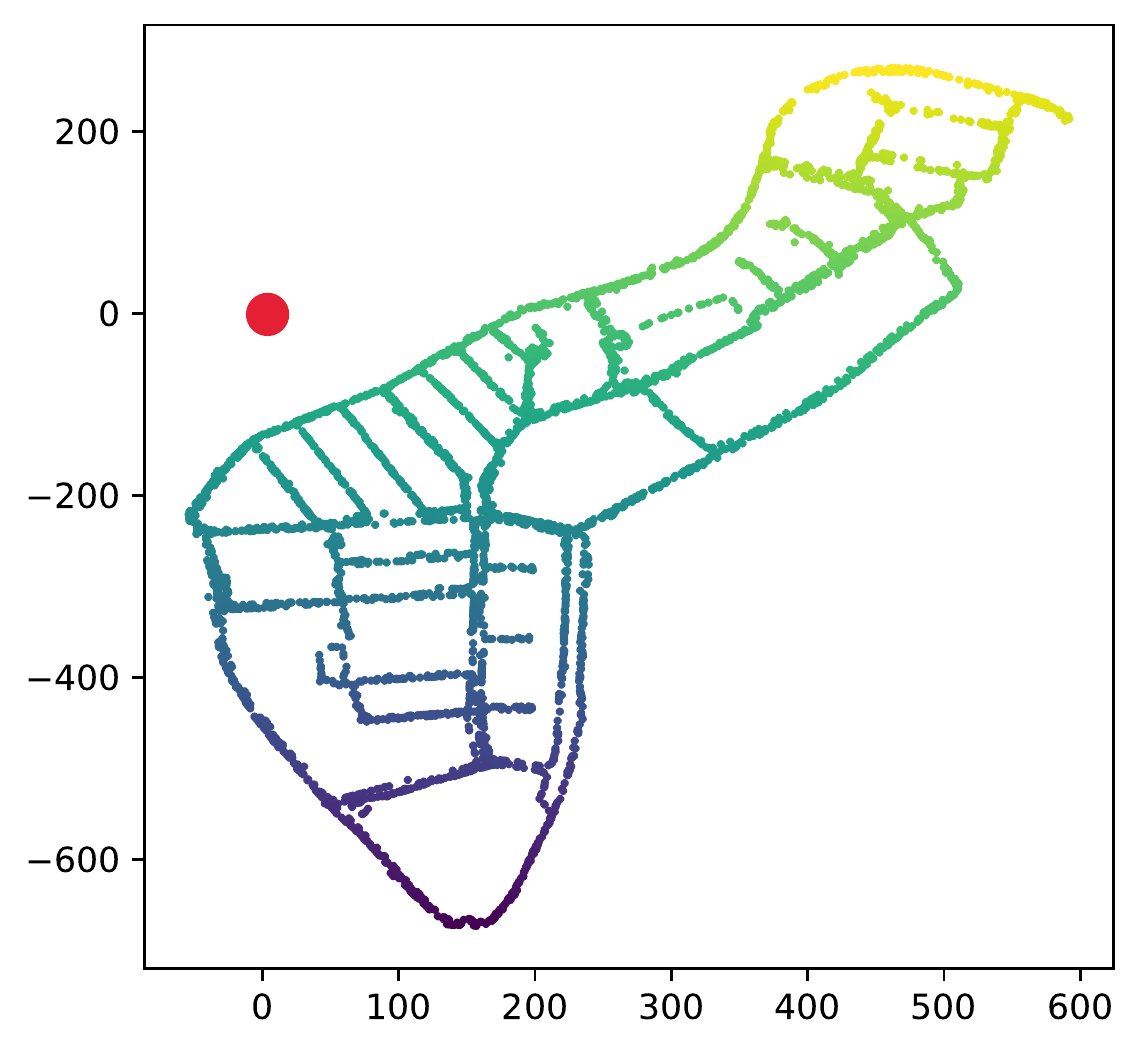}
        \vspace*{-6mm}
		\caption{}
	\end{subfigure}
\caption{a) Map of the measurement campaign in a residential area, where BS is marked with a red dot; b) Locations associated with the CSI measurements, which are coloured w.r.t. their \textrm{Y} coordinates~[m].}
\label{fig:map}
\end{figure}

We perform single-site positioning task on a real outdoor massive MIMO dataset
provided by IEEE Communications Theory Workshop data competition~\cite{CTW2020}.
The CSI measurements were taken in a residential
area of about 0.3 $\rm{km}^2$, as depicted in Fig.~\ref{fig:map} a).
The CSI data was collected by a channel sounder at a carrier frequency of 1.27 GHz, the  measurement details are given in~\cite{Gauger2020}. The data are made up of channel frequency responses between a moving single-antenna transmitter
and a fixed receiver which has a $8\times 8$ patch antenna array. The uplink OFDM channel has a bandwidth of 20 MHz with 1024
sub-carriers, of which 100 are used for guard bands, and $N$ = 924 are effective.
There are 8 antennas which were malfunctioning, so the effective antenna number is $B$ = 56.
Notice that the position of the effective antennas in the array is not provided.

The data contains a total of 4979 labeled CSI measurements with locations given by a GPS device, as shown in Fig.~\ref{fig:map} b).
To train the encoder $f_{\theta}(\mb H)$ and evaluate its positioning capability, we randomly split the data into a training and a test set. The ratio $\rho$ of test samples is set to be 10\% (with 498 samples) or 20\% (with 996 samples), similar to the settings in~\cite{Sobehy2021} and~\cite{Luc_LCOMM2021}.

More details of the DCNN-based CSI encoder depicted in Fig.~\ref{fig:DCNN} are given in Table \ref{tbl:tab1}. With those settings, the output size of the convolution part is $32\times 25 \times 25 = 20000$, which is then reduced by the projection head to a dimension of $R=32$. We implement the encoder and train it using the loss $\mc L$ via Pytorch on a desktop with a GeForce GTX 1660 Ti GPU. The default numbers of positives and negatives we used are $|\mathcal{P}_a|$ = 16 and $|\mathcal{N}_a|$ = 64. More details for training are also given in Table \ref{tbl:tab1}. It takes less than 10 minutes to learn a successful CSI encoder for the downstream kNN-based positioning task.
Finally, We feed all CSI samples in $\{ \mb H_i, \mb p_i\}_{{i=1\ldots I}}$ to the trained encoder, and obtain a CSI fingerprint database $\{ \mb z_i, \mb p_i\}_{{i=1\ldots I}}$.
\begin{table}[tbp]
\centering
\caption{Settings for the DCNN Encoder and Its Training.}
\label{tbl:tab1}	
\begin{tabular}{l|c}
\toprule
{Conv. layer 1}\!\!\!& {kernel size = 5, stride = 2, padding = 1} \\
\rowcolor{Gray}
{Conv. layer 2}      & {kernel size = 3, stride = 1, padding = 1} \\
{Conv. layer 3}      & {kernel size = 3, stride = 1, padding = 1} \\
\rowcolor{Gray}
{Conv. layer 4}      & {kernel size = 3, stride = 1, padding = 0} \\
{Feature space} & $R$ = 32  \\
\rowcolor{Gray}
{Optimizer}     & Adam    \\
{Learning rate} & 0.0005 (divided by 2 every 5 epochs)\\
\rowcolor{Gray}
{Weight decay}  & 0.0001 \\
{No. of epochs}  & 20    \\
\rowcolor{Gray}
{Batch size }    & $A$ = 32    \\
{Distance th.}  & $d_{\rm th}$ = 25\,m    \\
\rowcolor{Gray}
Temperature     & $\tau$ = 1.5 \\
\bottomrule
\end{tabular}		
\vspace{-0.2cm}	
\end{table}
\subsection{KNN-Based Positioning with the Learned CSI Similarity}
\begin{figure}[t!]
	\centering
	\begin{subfigure}{0.145\textwidth}
		\centering
		\includegraphics[width=\textwidth]{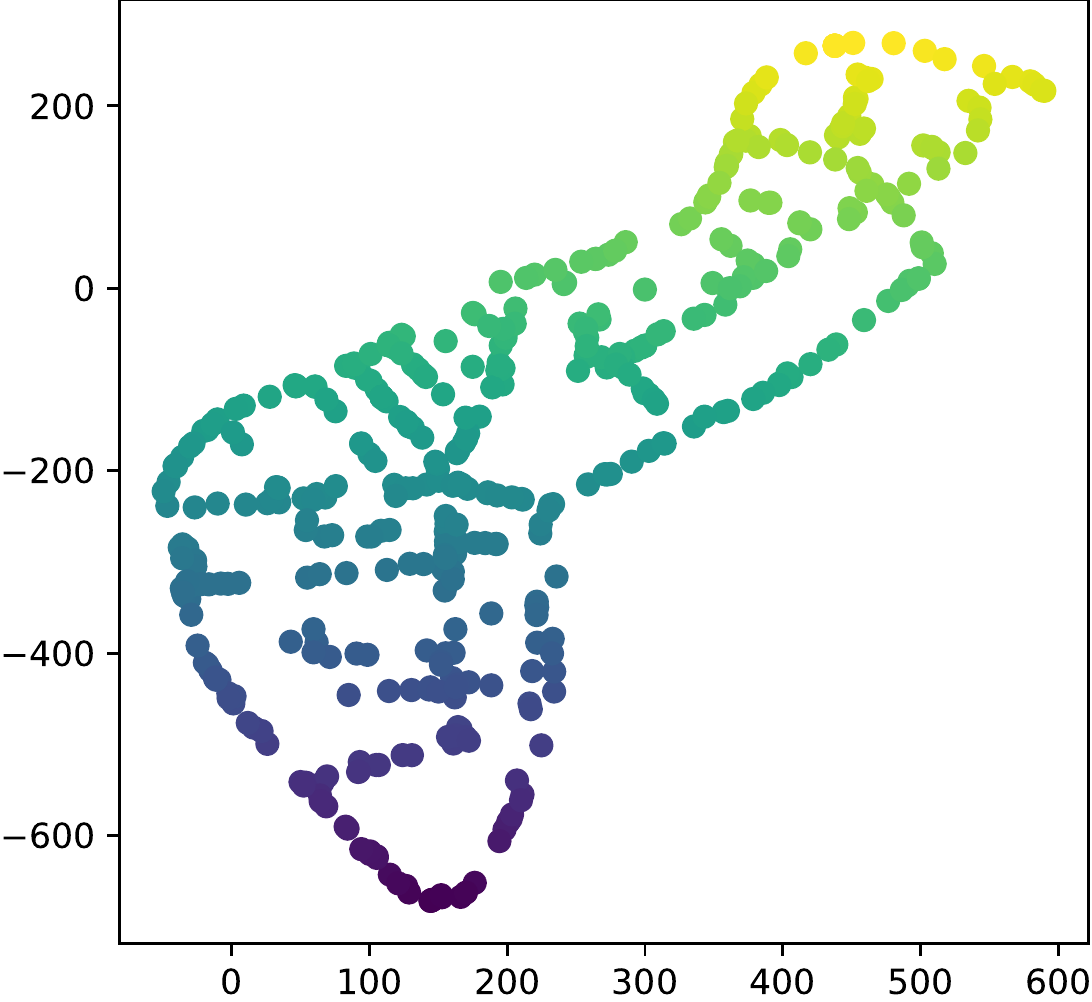}
		\caption{True.}
	\end{subfigure}
	\begin{subfigure}{0.145\textwidth}
		\centering
		\includegraphics[width=\textwidth]{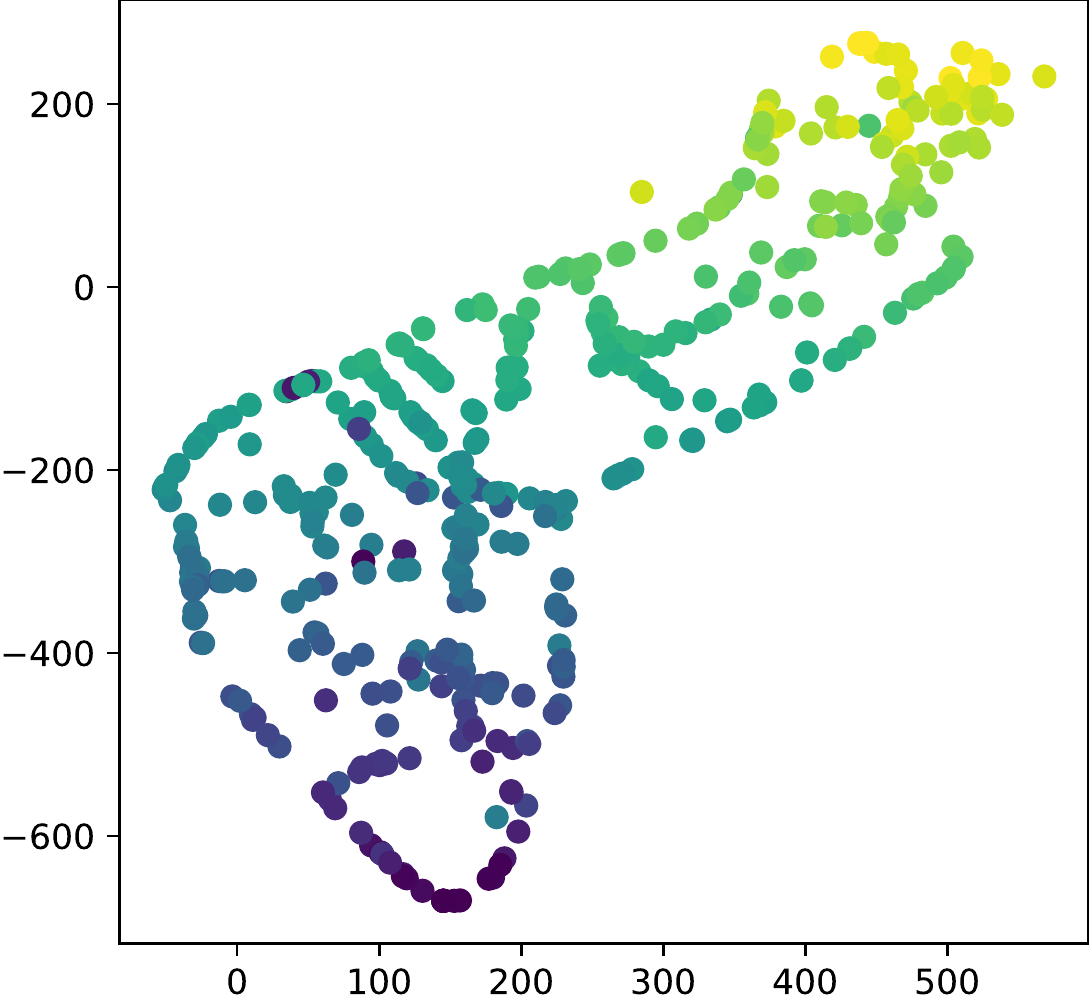}
		\caption{SupCon.}
	\end{subfigure}
    \begin{subfigure}{0.145\textwidth}
		\centering
		\includegraphics[width=\textwidth]{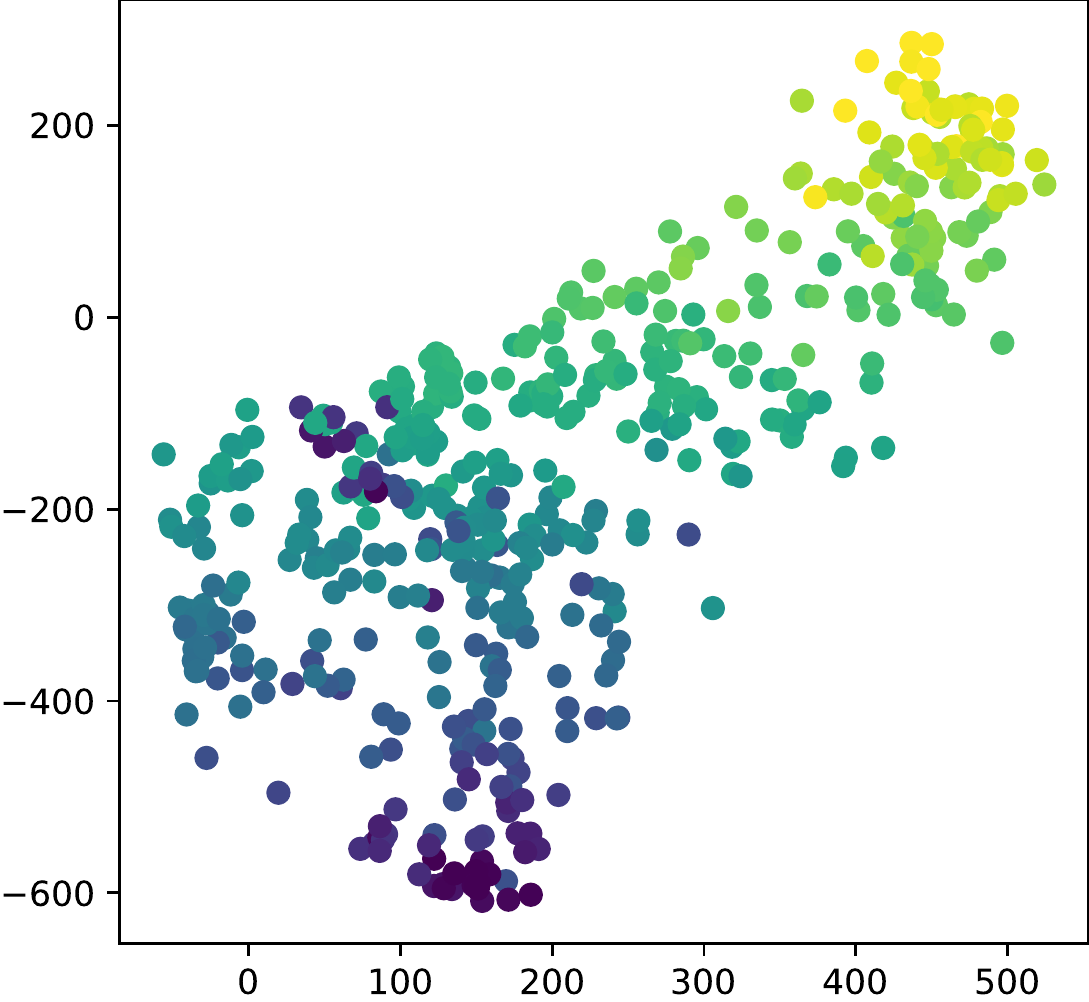}
		\caption{DM.}
	\end{subfigure}
	\begin{subfigure}{0.145\textwidth}
		\centering
		\includegraphics[width=\textwidth]{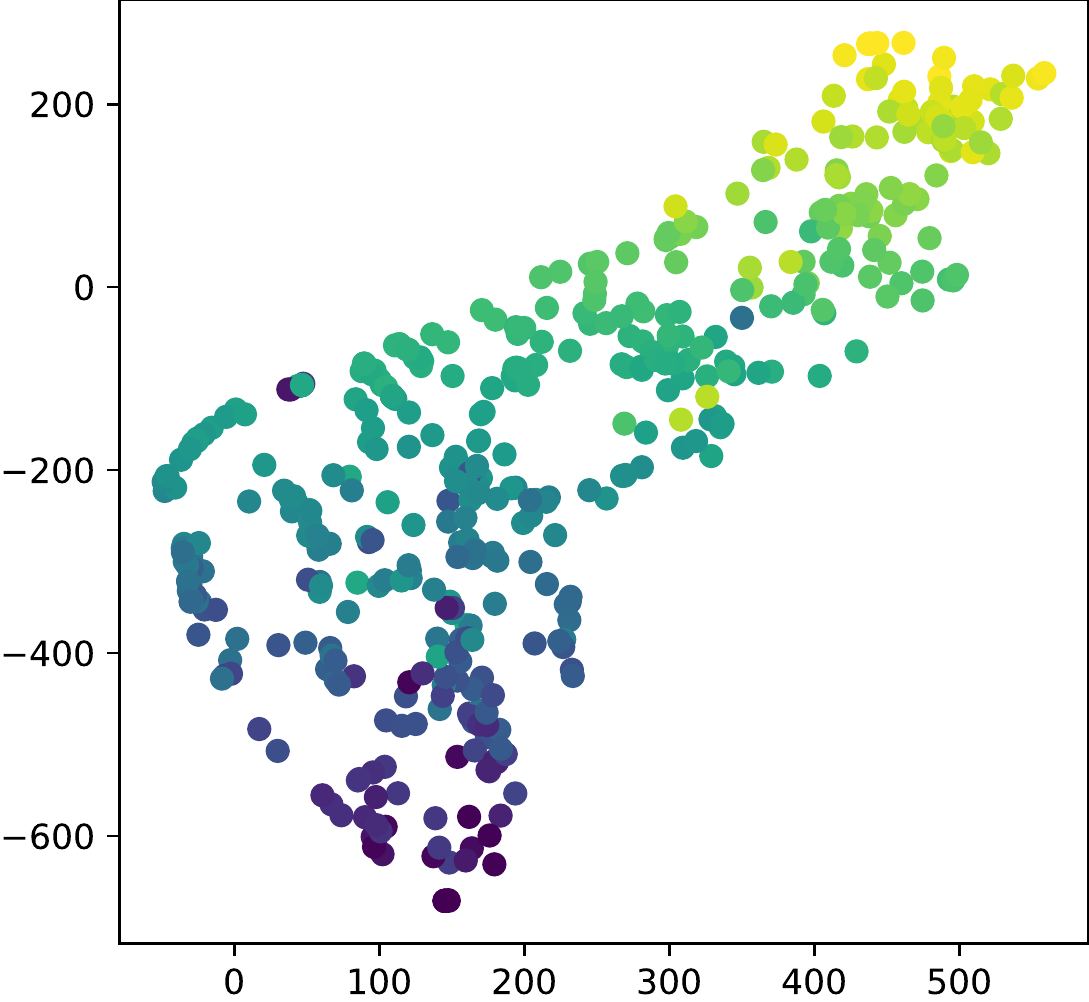}
		\caption{CMD.}
	\end{subfigure}
	\begin{subfigure}{0.145\textwidth}
		\centering
		\includegraphics[width=\textwidth]{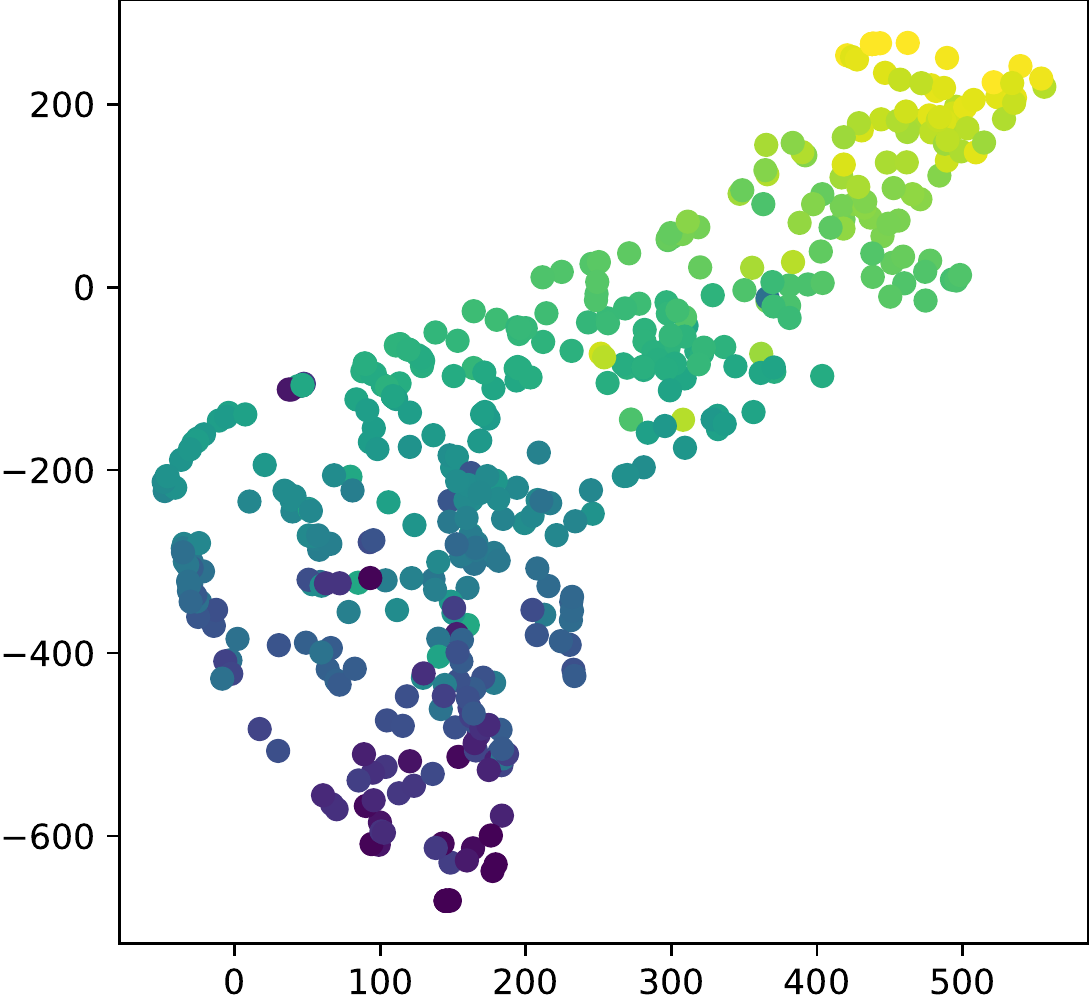}
		\caption{SVD.}
	\end{subfigure}
	\begin{subfigure}{0.15\textwidth}
		\centering
		\includegraphics[width=\textwidth]{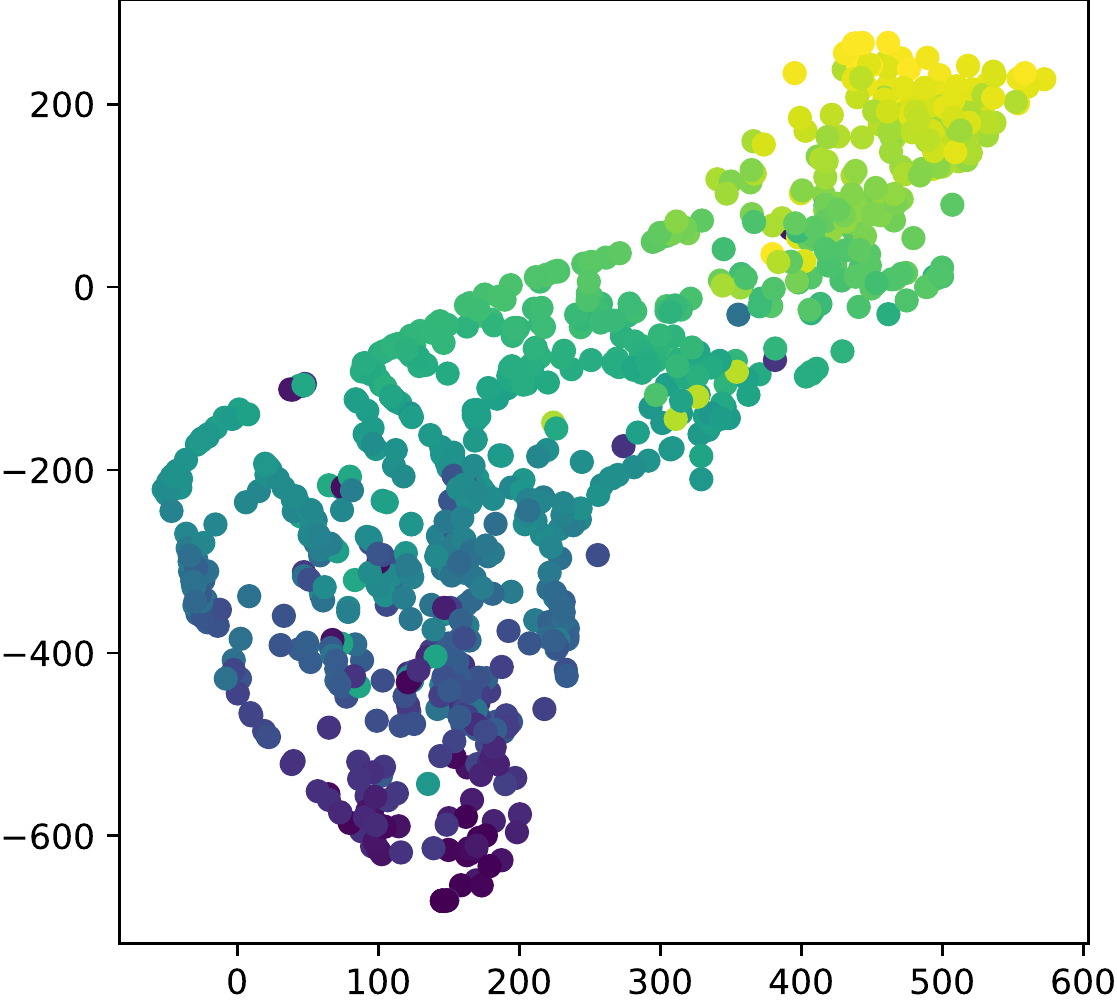}
		\caption{Magn.}
	\end{subfigure}
	\caption{ a) True positions of 498 test samples in the 2D XY-coordinates~[m];
    b) Predicted positions via the supervised contrastive learning~(SupCon) based similarity metric and kNN with \textrm{k} = 4 neighbors;
    c) Direct Mapping via a neural network in Fig.~\ref{fig:DCNN} with output size $R$ equals 2;
    d) kNN with correlation matrix distance~(CMD);
    e) kNN with SVD-based similarity;
    f) kNN with channel magnitude-based similarity.}
	\label{fig:scatters}
\end{figure}
For a test CSI $\mb H$ measured from an unknown location $\mb p$, we first compute its representation $\mb z = f_{\theta}(\mb H)$, and find its k-nearest neighbors~(denoted by $\mc K \subset \{1\ldots I\}$) in fingerprint database $\{ \mb z_i, \mb p_i\}_{{i=1\ldots I}}$, then the predicted position is given by
\begin{equation}\label{eq:wknn}
\begin{aligned}
\bar{\mb{p}} = \frac{\sum_{i\in \mc K} {\rm S}(\mb H_i, \mb H)\, \mb p_i }{\sum_{i\in \mc K} {\rm S}(\mb H_i, \mb H)}
=  \frac{\sum_{i\in \mc K} ||\mb z_i - \mb z||_2^{-1} \mb p_i }{\sum_{i\in \mc K} ||\mb z_i - \mb z||_2^{-1}},
\end{aligned}
\end{equation}
and the positioning error is evaluated as $||\mb{p}-\bar{\mb{p}}||_2$.
Using 498 CSI measurements for test (with $I=4481$ for training), their predicted locations are plotted in Fig.~\ref{fig:scatters} b).
As compared with the true positions shown in Fig.~\ref{fig:scatters} a), we achieve a mean positioning error of 35.2\,m, which is better than the lowest errors reported in~\cite{Luc_LCOMM2021},\cite{GarauBurguera2020} and~\cite{Sobehy2021} as 42\,m, 40\,m and 37\,m respectively. In~\cite{Luc_LCOMM2021}, its best result was achieved by converting
the Nadaraya-Watson estimator to a three-layer neural network and using a SVD-based similarity metric. In~\cite{GarauBurguera2020}, its best performance was given by kNN with a sophisticated similarity metric based on signal subspace, whose computation complexity is rather high. While in~\cite{Sobehy2021}, the 37\,m mean error is achieved via ensemble learning with multi-layer perceptron neural networks~(MLP NN), and hand-crafted CSI features extracted from channel magnitude information via polynomial regression and Fourier fitting. Compared with these methods, our proposed positioning pipeline not only has better positioning performance, but also is easier to be implemented.
\subsection{Comparison with other CSI Similarity Metrics}
To gain more insights about the learned similarity metric, in our experiments, we compare it with three other CSI similarity metrics, i.e. CMD, SVD-based, and channel magnitude-based, by replacing ${\rm S}(\mb H_i, \mb H)$ in~\eqref{eq:wknn}.
The CMD similarity~\cite{Wu_TWC_2021} is
\begin{align}\label{eq:cmd}
 {\rm S}_{\rm CMD}(\mb H_i, \mb H) = \frac{{\rm{Tr}}\{ (\mb H_i^{} \mb H_i^{\dagger})  (\mb H \mb H^{\dagger}) \}}{||(\mb H_i^{} \mb H_i^{\dagger})||_{\rm F} \, ||(\mb H \mb H^{\dagger})||_{\rm F}},
\end{align}
with $\rm{Tr\{\cdot\}}$ denotes matrix trace and $||\cdot||_{\rm F}$ the Frobenius norm. The SVD-based similarity~\cite{Luc_LCOMM2021} is defined as
\begin{align}\label{eq:chordal}
 {\rm S}_{\rm SVD}(\mb H_i, \mb H) = |\mb v^{\dagger}_i \mb v^{}|,
\end{align}
where $\mb v$ is the left singular vector w.r.t. the largest singular value of $\mb H$.
The magnitude-based similarity is defined as
\begin{align}\label{eq:Magnitude}
 {\rm S}_{\rm Magn}(\mb H_i, \mb H) = {\left[\sum\nolimits_{b} \sum\nolimits_{n} \left(|\mb H_i|_{b,n} - |\mb H|_{b,n}\right)^2\right]^{-1}},
\end{align}
where $|\mb H|_{b,n}$ is the ($b,n$)\,th element of the channel magnitude matrix $|\mb H|$ of $\mb H$.
%

\begin{figure}[t!]
\centering
\begin{tikzpicture}
  \node (img1)  {\includegraphics[width=0.32\textwidth]{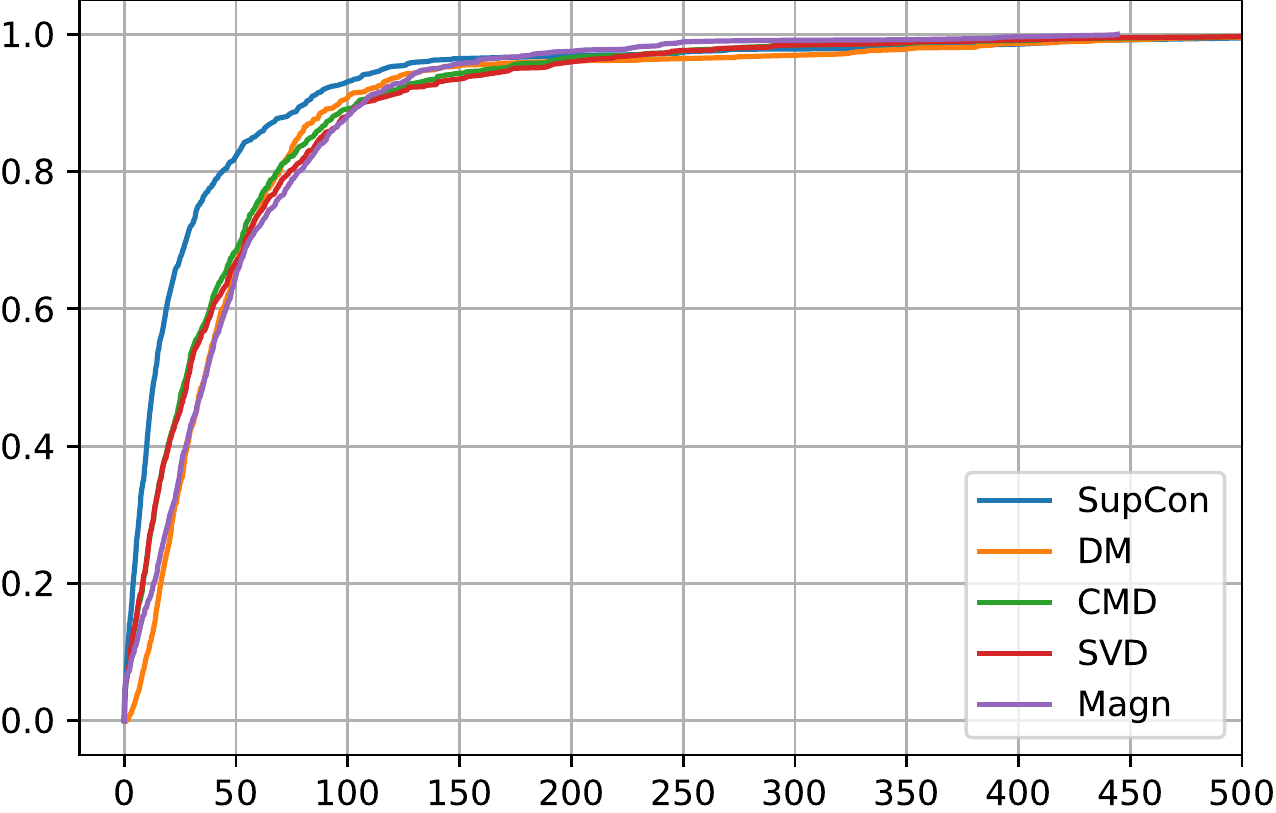}};
  \node[below=of img1, node distance=0cm, yshift=1.2cm] {\scriptsize{positioning error~[m]}};
  \node[left=of img1, node distance=0cm, rotate=90, anchor=center,yshift=-0.7cm] {\scriptsize {CDF}};
\end{tikzpicture}
\vspace{-0.2cm}	
\caption{CDF of positioning errors for kNN with different CSI similarity metrics~(i.e. SupCon, CMD, SVD-based and magnitude-based), and direct mapping~(DM).
Number of test samples is 498. }
\label{fig:CDF}
\end{figure}

For the 498 test samples, their positioning results are shown as in~Fig.~\ref{fig:scatters} d), e), f). Compared with our CSI similarity, CMD has a mean positioning error of 48.3\,m, SVD-based similarity has one of 50.8\,m, while it is 56.6\,m for channel magnitude-based, as summarized in Table~\ref{tbl:tab2}. Here, we have set k = 4 neighbors for kNN as it produced the best results for all similarity metrics in our experiments.
The cumulative distribution functions~(CDFs) of their positioning errors, together with that for our method are given in Fig.~\ref{fig:CDF}. The error distribution has a long tail because the CSI samples from the northeast and southernmost regions exhibit very low SNRs.  With our method, 66\% of the test samples have a positioning error smaller than 25\,m, while other methods have less than 47\% of the test samples achieve this goal.
\subsection{Comparison with Direct Mapping}
The object of direct mapping~(DM)~\cite{GLOBECOM2020_DNN,Bast_VTC_2020,EuCNC2021} is to train a CSI encoder such that CSI is mapped to a point in the 2D (or 3D) geographic space directly.
For DM, we use the same network architecture as in Fig.~\ref{fig:DCNN}, with an output dimension of $R=2$, while other network parameters are the same as for contrastive CSI feature learning. To train the DM encoder $\mb z = f_{\theta,\rm{DM}} (\mb H)$, a loss function  $\mc L_{\rm DM} =  \frac{1}{A} \sum_{a \in \mc A} ||\mb z_a -\mb p_a||_2$ is used. Different from Table \ref{tbl:tab1}, we use an initial learning rate of 0.01 and train the DM encoder for 100 epochs. For a new CSI $\mb H$, the predicted position is then given by $\bar{\mb p} = f_{\theta,\rm{DM}} (\mb H)$.

\begin{figure}[t!]
\centering
\begin{tikzpicture}
  \node (img1)  {\includegraphics[width=0.32\textwidth]{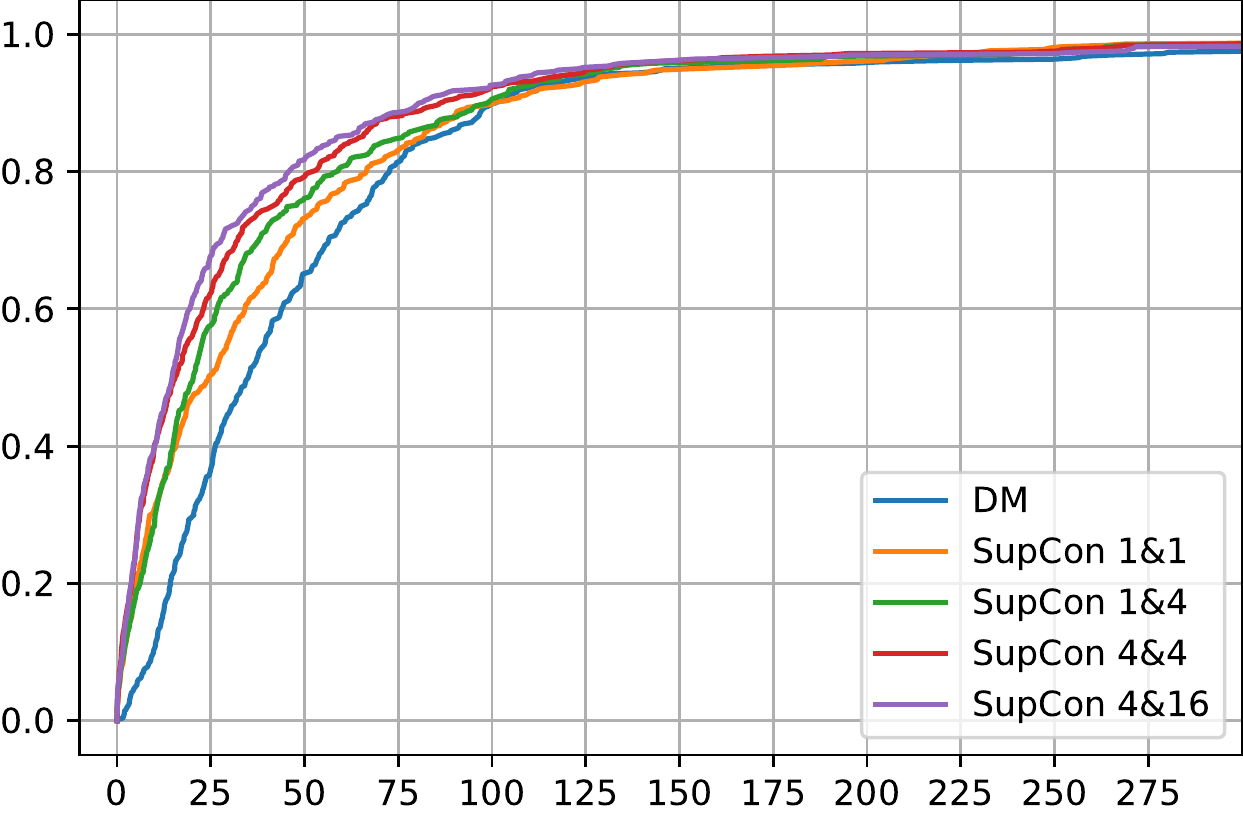}};
  \node[below=of img1, node distance=0cm, yshift=1.2cm] {\scriptsize{positioning error~[m]}};
  \node[left=of img1, node distance=0cm, rotate=90, anchor=center,yshift=-0.7cm] {\scriptsize {CDF}};
\end{tikzpicture}
\vspace{-0.2cm}	
\caption{CDF of positioning errors on the test set for direct mapping~(DM), and kNN based on SupCon CSI similarities with different numbers ($|\mathcal{P}_a|$ \& $|\mathcal{N}_a|$) of positives and negatives. Number of test samples is 498. }
\label{fig:CDF2}
\end{figure}

The positioning results using DM is shown in Fig.~\ref{fig:scatters} c). It exhibits a mean error of 54.3\,m, which is close to kNN with CMD, SVD-based and magnitude-based similarities, but much worse than our contrastive learning-based approach. It indicates that the proposed DCNN-based CSI encoder network do have the capability to learn a inverse mapping from high-dimensional CSI space to the geographic space for the black box model $\mb H = \mc G (\mb p,\mc X)$. However, due to the sparsity of the training data and over fitting problem, the learned DM encoder has high generalization errors on the test set.
Instead, the proposed  contrastive learning method tries to learn an intermediate CSI feature space where CSI's representations exhibit similar neighbour relations as in the geographic space, supervised by the relative distance information of positives and negatives.
As shown in Fig.~\ref{fig:CDF2}, with only one positive and one negative, a mean positioning error of 45.1\,m can be achieved and the performance improves  as both the number of positives $|\mathcal{P}_a|$ and number of negatives $|\mathcal{N}_a|$ increase.

\begin{table}[t!]
\centering
\caption{Mean Positioning Errors [m].}
\label{tbl:tab2}	
\begin{tabular}{@{}lccccc@{}}
			\toprule
			{}\!\!\!& DM & SupCon\!\!\! & CMD\! & SVD \!\!\!& Magn. \\
			\midrule
			{$\rho$ = 10 \%} & 54.3 & \textbf{35.2} & 48.3 & 50.8 & 56.6 \\
			{$\rho$ = 20 \%} & 55.5 & \textbf{36.4} & 48.7 & 51.4 & 57.3 \\
			\bottomrule
\end{tabular}		
\vspace{-0.2cm}	
\end{table}

\section{Conclusions}
We have proposed a novel contrastive CSI representation learning method for massive MIMO positioning. A contrastive loss function has been designed, involving multiple positive and negative  samples. A versatile DCNN-based CSI encoder we designed has been trained using this loss  to learn robust CSI representations for fingerprinting-based positioning task. We have conducted extensive experiments on a real-world massive MIMO dataset measured in a complex outdoor environment. The results show that the learned CSI similarity metric improves the positioning accuracy significantly compared with other known methods. No specific knowledge of the array structure and array calibration are needed for the proposed positioning pipeline.
Instead of using a static kNN estimator in the final step, one can also consider using a trainable regressor as in~\cite{Luc_LCOMM2021} to further improve the positioning accuracy by end-to-end fine-turning. Another promising research direction is to combine the learned similarity metric with the semi-supervised channel charting methods~\cite{IWCMC2021,ICCC2021}, considering  neighborhood relationships in the feature space is generally non-linear.
\bibliographystyle{IEEEtran}
\bibliography{SupConCSI}
\end{document}